\documentclass{aa}
\usepackage{graphicx}
\usepackage{txfonts}
\usepackage{lscape}
\usepackage{natbib}

\newcommand{\ctss}{\mbox{counts s}^{-1}}

\def\ergscm2{\rm erg\,cm^{-2}\,s^{-1}}

\def\xss{XSS\,J1227}

\begin{document}

\title{X-ray follow-ups of 
XSS\,J12270-4859: a low-mass X-ray binary with gamma-ray Fermi-LAT 
association\thanks{Based on observations obtained 
with \emph{XMM-Newton}, an 
ESA science mission with instruments and contributions 
directly funded by  ESA Member States and NASA, with 
\emph{Swift},  a NASA science mission with Italian participation, 
with  \emph{Rossi-XTE} , a NASA science mission, and 
with \emph{Fermi} a 
NASA  mission with contributions from France, Germany, Italy, Japan, 
Sweden, and U.S.A., and with \emph{AGILE}, an Italian Space Agency mission with participation 
of the Italian Institute of Astrophysics and the Italian Institute of Nuclear
Physics.}}

\author{D.~de Martino\inst{1}
\and
T. Belloni\inst{2}
\and
M.~Falanga\inst{3}
\and
A.~Papitto\inst{4}
\and
S.~Motta\inst{2}
\and
A.~Pellizzoni\inst{5}
\and
Y.~Evangelista\inst{6}
\and
G.~Piano\inst{6}
\and
N.~Masetti\inst{7}
\and
J.-M.~Bonnet-Bidaud\inst{8}
\and
M.~Mouchet\inst{9}
\and
K.~Mukai\inst{10}
\and
A.~Possenti\inst{5}
}


\institute{
INAF - Osservatorio Astronomico di Capodimonte, salita Moiariello 16, I-80131 Napoli, Italy \\
\email{demartino@oacn.inaf.it}
\and
INAF-Osservatorio Astronomico di Brera, Via E Bianchi 46, I-23807 Merate (LC) Italy\\
\email{tomaso.belloni@brera.inaf.it,sara.motta@brera.inaf.it}
\and
International Space Science Institute (ISSI), Hallerstrasse 6,
CH-3012 Bern, Switzerland \\
\email{mfalanga@issibern.ch}
\and
Institut de Ci\'encie de l'Espai (IEEC-CSIC) Campus UAB, Fac. de 
Ci\'encies, Torre C5 E-08193 Barcelona, Spain\\
\email{papitto@ice.csic.es} 
\and
INAF - Osservatorio Astronomico di Cagliari, loc. Poggio dei Pini, strada 
54, I-09012, Capoterra (CA), Italy\\
\email{apellizz@oa-cagliari.inaf.it} 
\and
INAF - Istituto di Astrofisica Spaziale e Planetologia Spaziali, Via Fosso del 
Cavaliere 100, Roma, I-00133 ITALY \\
\email{yuri.evangelista@iasf-roma.inaf.it, giovanni.piano@iasf-roma.inaf.it}
\and
INAF - Istituto di di Astrofisica Spaziale , Via Gobetti 101, I-40129, Bologna, Italy \\
\email{nicola.masetti@iasfbo.inaf.it}
\and
CEA Saclay,  DSM/Irfu/Service d'Astrophysique, F-91191 
Gif-sur-Yvette, France \\
\email{bonnetbidaud]@cea.fr}
\and
Laboratoire APC, Universit\'{e} Denis Diderot, 10 rue Alice Domon et L\'{e}onie Duquet, F-75005
Paris, France and LUTH, Observatoire de Paris, Section de Meudon, 5 place Jules Janssen, F-92195
Meudon, France \\
\email{martine.mouchet@obspm.fr}
\and
 CRESST and X-Ray Astrophysics Laboratory, NASA Goddard Space Flight 
 Center, Greenbelt, MD 20771,
 USA and Department of Physics, University of Maryland, Baltimore County, 
 1000 Hilltop Circle,Baltimore, MD 21250, USA \\
\email{koji.mukai@nasa.gov}
}

\date{Received September 18, 2012; accepted December 2, 2012}

\abstract 
{XSS\,J1227.0-4859 is a peculiar, hard X-ray source recently positionally associated to the 
\emph{Fermi}-LAT source 
 1FGL\,J1227.9-4852/2FGL\,J1227.7-4853. Multi-wavelength observations have added information on this 
source, 
indicating a low-luminosity low-mass X-ray binary (LMXB), but its nature is still unclear.}
{To progress in our understanding, we present new X-ray data from a monitoring campaign performed in 2011 
with the \emph{XMM-Newton},   \emph{RXTE}, and \emph{Swift} satellites and combine them with new gamma-ray 
data from the \emph{Fermi} and 
\emph{AGILE} satellites. We complement the study with simultaneous near-UV photometry from \emph{XMM-Newton} and
with previous UV/optical and near-IR data.} 
{We  analysed the temporal characteristics in the X-rays, near-UV, and gamma rays and studied the 
broad-band spectral energy distribution from radio to gamma rays.}
{The X-ray history of XSS\,J1227 over 7\,yr shows  a persistent and rather 
stable low-luminosity ($\rm 6\times 10^{33}\,d_{1\,kpc}^2 erg\,s^{-1}$) source, with  flares and dips being 
 peculiar and permanent characteristics. The associated \emph{Fermi}-LAT 
source 2FGL\,J1227.7-4853 is 
also stable over an overlapping period of 4.7\,yr. 
Searches  for X-ray fast pulsations down to msec give upper limits to 
pulse  fractional amplitudes of $15-25\%$ 
that do not rule out a fast spinning pulsar. 
The combined UV/optical/near-IR spectrum reveals  a hot component at $\sim$13\,kK and a cool one 
at $\sim$4.6\,kK. The latter would suggest a late-type K2-K5 companion star,  a distance range of 
1.4--3.6\,kpc, and an orbital period of 7--9\,h. A near-UV variability ($\gtrsim$6\,h) also suggests a longer
orbital period than previously estimated.}
{The analysis shows that the X-ray and UV/optical/near-IR emissions are more compatible with an accretion-powered 
compact object than with  a rotational  powered pulsar. 
The X-ray to UV bolometric luminosity ratio could be consistent with a binary hosting 
a neutron star, but the uncertainties in the radio data may also allow an LMXB black hole with a 
compact 
jet.In this case, it would be the  first  associated with a high-energy gamma-ray source.}


\titlerunning{Follow-ups of XSS~J12270-4859}

\maketitle

\section{Introduction}

XSS~J12270-4859 (henceforth XSS\,J1227) is a peculiar and enigmatic hard X-ray source.
An early proposal that it could be a cataclysmic variable (CV) hosting a 
magnetic white dwarf \citep{masetti06,Butters08} was disregarded with independent observations by 
\citet{Pretorius09}, \citet[]{Saitou}, and \citet[][in the following dM10]{deMartino10}. These observations instead 
suggested  a low-mass 
X-ray  binary (LMXB) nature. A putative 4.32\,h orbital period was 
also claimed from optical and possibly near IR photometry by dM10, but not confirmed by 
further  near-IR (nIR) data \citep{Saitou11}. In dM10 we discovered this source to be positionally associated to 
the bright unidentified gamma-ray \emph{Fermi}-LAT source, 1FGL\,J1227.9-4852, detected up to 10\,GeV.
1FGL\,J1227.9-4852 is also reported in the  $\rm 2^{nd}$  \emph{Fermi}-LAT catalogue as 
2FGL\,J1227.7-4853 \citep{Nolan12} (henceforth 2FGL\,J1227). Based on this association, we found that the 
gamma-ray emission is a non-negligible fraction of the X-ray luminosity ($\rm 
L_{0.1-100\,GeV}/L_{0.2-100\,keV} \sim$ 0.8) and if the X-ray and gamma-ray emissions are linked, 
the peak energy should be between 1--100\,MeV.

\noindent Due to its peculiar flaring and dipping behaviour,  XSS\,J1227 was proposed to share some 
similarities, though with rather different timescales and energetics, with type-II burst sources and
to represent an unusual low-luminosity ($\rm L_{X} \sim 6\times 10^{33}\,erg\,s^{-1}$ at 1\,kpc) LMXB  (dM10).

The nature of the compact object has been further 
investigated in  other wavelength domains \citep{Saitou11,Hill}. 
 From radio searches conducted by  \citet{Hill},  three sources 
within the  \emph{Fermi}-LAT 99$\%$ error box were detected, with only one having an obvious  X-ray 
counterpart. This 
faint  radio source is located at  the position of XSS\,J1227 with  flux densities  of 0.18\,mJy at 5.5\,GHz and of 
0.14mJy at 9\,GHz. It was not detected at lower frequencies (640\,MHz, 240\,MHz, and at 1400\,MHz) in follow-up GMRT 
observations. For a power-law spectrum 
$\rm S_{\nu} \sim \nu^{\alpha}$, the radio data gave 
a fairly unconstrained power-law index $\alpha$=-0.5$\pm$0.6. Also, \cite{Hill} did not detect fast radio pulses down 
to  milliseconds  (msec) in subsequent  Parkes observations.
Nonetheless, the association  with the  \emph{Fermi}-LAT source favoured XSS\,J1227 as a msec pulsar (MSP) 
binary  \citep{Hill}, similar to the first discovered rotational-powered MSP in a quiescent LMXB PSR\,J1023+0038 
\citep{Tam}. 
The number of MSP  binaries detected in gamma-rays by   \emph{Fermi}-LAT 
has rapidly increased in the last year with the newly discovered systems  
PSR\,J1231-1411, PSR\,J0614-3329, and  PSR\,J2214+3000 \citep{Ransom11},  PSR\,J1816+4510  
\citep{Kaplan12},
PSR\,J0101-6422,  PSR\,J1514-4946, and PSR\,J1902-5105  \citep{Kerr12}. A few are found to
host so-called black widow pulsars such as PSR\,J2051-0827\citep{Wu12} and J\,2339+0533 \citep{Romani11,Kong12}. 
A number of them are detected as faint X-ray sources with  X-ray luminosities typical of radio MSPs (e.g. 
\citet{Ransom11,Wu12}). Hence, it is expected that the \emph{Fermi} satellite will discover most of the local population 
of MSPs, thus allowing constraint of the emission mechanisms and the binary evolution of their progenitors.

On the other hand, from the detection of simultaneous X-ray and nIR flares and the similarity of broad-band 
spectral energy distribution (SED) at low energies with black hole (BH) candidates such as GRS\,1915+105, XTE\,J1118+480 and 
Cyg\,X-3,   
\cite{Saitou11} proposed that XSS\,J1227 is reminiscent of 
microquasars  with a synchrotron  jet.  However, broad-band SEDs,  if due to synchrotron cooling in optically thin 
regime,  are expected to show a break at $\sim$500\,keV, making them undetectable in the high-energy 
gamma-rays. 
 The  recent  \emph{Fermi} and \emph{AGILE} detections of high-mass X-ray binaries (HMXBs) 
\citep{abdo09c,Tavani09,Ackermann12,Hadasch}, with some of them still thought to host neutron stars 
(NS) or BH 
\citep{Hadasch,Papitto12,Torres12} may open the possibility of detecting faint gamma-ray sources
in different binary types \citep{Hadasch}.

\begin{figure*}[t!]
\resizebox{\hsize}{!}{\includegraphics{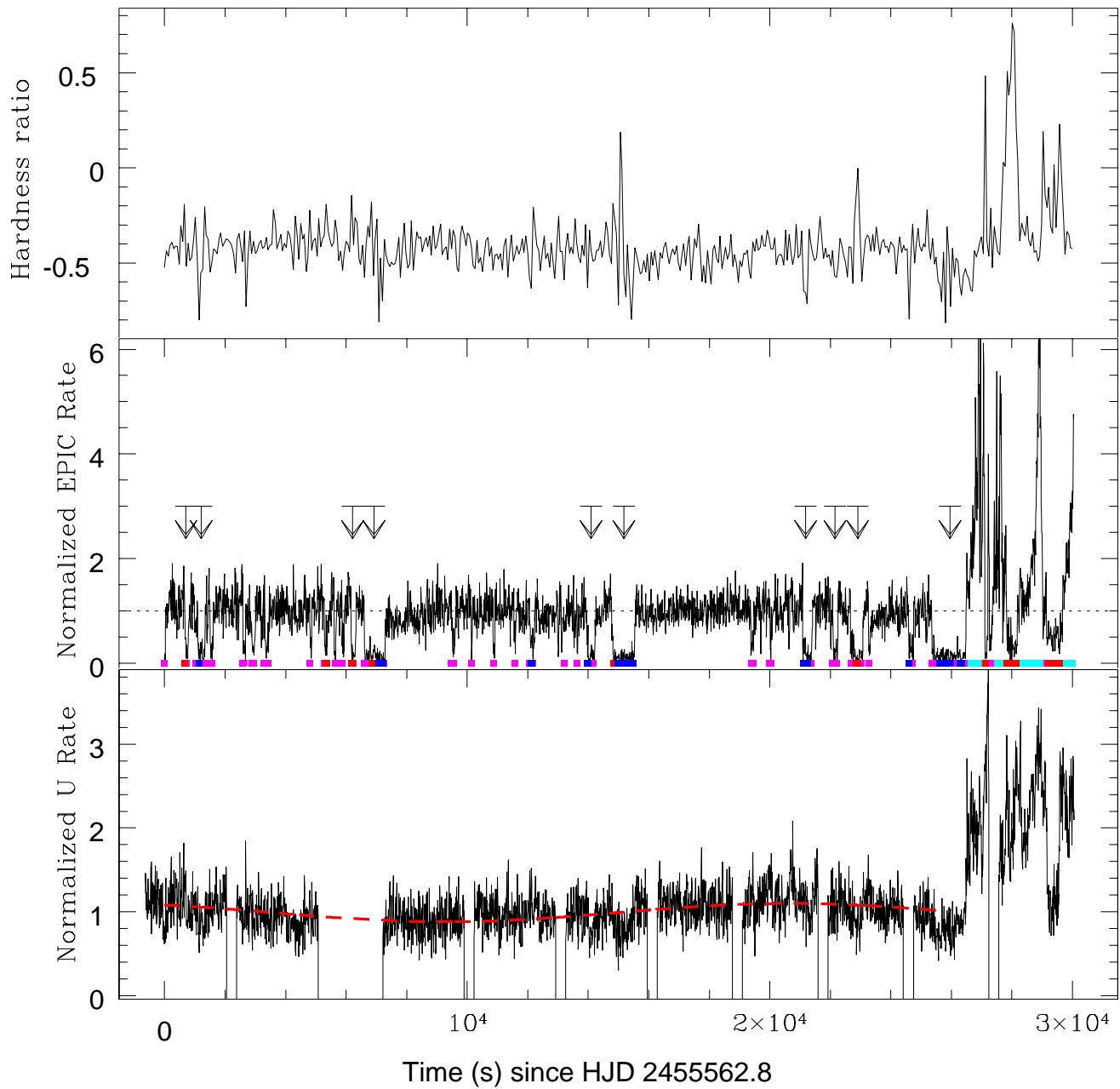}}
\caption{{\it Bottom panel:} Net U band light curve with a 10sec binning 
time. Count rates are normalized 
to the average persistent quiescence value. A sinusoidal fit with a period of 6.4\,h is also reported. Gaps 
are due
to the OM fast acquisition mode.
{\it Middle panel:} The EPIC-pn net light curve in the  0.3-10\,keV range with 
a 10\,s binning time.  The count rate is normalized to the average quiescent value. Flares are marked 
in cyan,  dips with no spectral change in magenta,  dips showing a softening are marked in blue, 
while hard dips are marked 
in red.
Arrows mark ten long dips used to study the CCF and spectra. 
{\it Top panel:} The hardness  ratio between the  0.3-2\,keV and 2-10\,keV bands with binning time of 60\,s. 
See electronic edition for colour version. }
\label{HR_time_PN}
\end{figure*}

To further advance  our understanding of this intriguing object, we started a X-ray 
monitoring campaign in 2011 to infer whether this source 
undergoes changes in luminosity and to detect possible fast (msec) pulsations in the 
X-ray band. We were also granted  \emph{AGILE}/GRID data acquired since the
beginning of operations in 2007  to detect for the first time this source in the soft gamma rays down 
to 50\,MeV to obtain a broad-band gamma-ray coverage combining the GRID and LAT data.
With these  aims, we present the X-ray 
observations acquired along one year with 
\emph{Swift} and \emph{Rossi-XTE} (\emph{RXTE})  and the  fast timing X-ray data acquired with 
\emph{XMM-Newton}.  In Sect.\,2 we report the observations and data reduction; in Sect.\,3 
we provide analysis of the X-ray
light curves, including the long-term comparison with the gamma-ray \emph{Fermi}-LAT curve and the 
search for fast msec pulses; in 
Sect.\,4 we examine the spectral properties over a wide energy range, and in Sect.\,5, we discuss the 
possible nature of this object.

\section{Observations and data reduction}

The new observations acquired with \emph{XMM-Newton}, \emph{RXTE},
and \emph{Swift} as well as the time coverage of \emph{AGILE}/GRID are summarised in 
Table\,\ref{observ}.


\subsection{The \emph{XMM-Newton} observation}

A 30\,ks pointing with  \emph{XMM-Newton} (OBSID: 656780901) was carried out on Jan.1, 2011. In 
contrast to our past observation in 2009 (dM10), the EPIC-pn camera \citep{struder01} was 
operated in the  timing read-out mode that reaches a resolution of 0.03\,ms.  The EPIC-MOS cameras 
\citep{turner01}  were instead operated in imaging full-window mode using the thin filters. The 
Optical Monitor (OM) \citep{mason01} was operated in fast-window mode using the U (3000--3800 \AA) 
filter throughout the observation
(see Table\,\ref{observ}).  The data were processed  using the standard reduction 
pipelines and analysed with the SAS  10.0 package using the latest calibration files. 
The photon arrival times from EPIC cameras and OM were reported to the solar system barycentre using 
the  nominal position of XSS\,J1227 \citep{masetti06}. 
In the timing-mode read-out used for the EPIC-pn exposure, the spatial information is lost because
imaging is made only in one dimension and the data from a predefined area on one CCD chip are collapsed 
into a one-dimensional  row to be read out at high speed. 
The EPIC-pn  events from the source and background were extracted from RAWX=24--52px and from 
RAWX=3-15px, respectively.
For the EPIC-MOS cameras, we instead extracted events using a  circular region of 37'' aperture radius centred on the source 
and using a background region located on the  same CCD chip. In order to improve the S/N ratio, 
we filtered the data by selecting pattern pixel events up to 
double with zero-quality flag for the EPIC-pn data,  and up to quadruple pixel events for the 
EPIC-MOS data. 
The average  background level of the EPIC cameras was low during the whole observation. 
The background-subtracted OM-U light curve was obtained with a binning 
time of 10\,s.

\subsection{The \emph{RXTE} observations}

In order to infer the mid- to long-term behaviour of XSS\,J1227, we initiated a 
 monitoring with \emph{RXTE} \citep{bradt_etal93} (Prog.ID: 96309) that started in Jan. 2011  and
ended in Dec. 2011. A total of 44 \emph{RXTE}/PCA pointings were performed with a cadence of 
about one week and typical 
exposure times of 2\,ks (see Table\,\ref{observ}). 
The \emph{RXTE}/PCA standard data products for each observations were obtained from the \emph{RXTE} Guest Observer 
facility. Barycentric corrections were applied to the background-subtracted light curves.  
XSS\,J1227 was typically found at a net count rate 
of $\sim$0.3-0.5\,cts\,s$^{-1}$/PCU in the 9-20\,keV. 
Due to the poor statistics above 10\,keV, the timing analysis 
is restricted to the 2-9\,keV range, also using the soft 2-4\,keV and hard 4-9\,keV bands. 
The source spectra were instead analysed up to 30\,keV in each observation.

\subsection{The \emph{Swift} observations}

The monitoring of XSS\,J1227 with \emph{Swift}/XRT (Prog.ID:41135) started in Mar. 2011 and 
ended in Sept. 2011. A 
total of 
19 snapshots were performed with typical exposure times of 800\,s. A denser coverage was performed on Mar.\,23 and 
on Sept.\,19 (see Table\,\ref{observ}). We used the \emph{Swift}/XRT data products generator at the University of 
Leicester \citep{Evans09} to build background-subtracted light curves in the 0.3-10\,keV, 
0.3-1.5\,keV, and 1.5-10\,keV bands 
and  spectra in the 0.3-10\,keV range. We used photon counting-mode data only. Barycentric 
corrections 
were applied to the  extracted light curves. 

\subsection{The \emph{AGILE} data}

\emph{AGILE} (Astrorivelatore Gamma ad Immagini LEggero) \citep{Tavani09a}
consists of a large field of view ($\sim60^{o}$) gamma-ray imager,  
GRID, sensitive in the energy range  30\,MeV--50\,GeV, and a co-axial hard
X-ray detector (Super-AGILE) for imaging in the 18--60\,keV range \citep{Feroci}.
Since beginning operation in Jul. 2007, \emph{AGILE} has worked in pointing mode, 
but science operations were reconfigured following a malfunction of the rotation wheel in 
mid-Oct. 2009. Since then,  the satellite has been  operating in ''spinning observing mode", with 
the solar panels  pointing at  the Sun and the instrument axis sweeping the sky with an angular speed 
$\rm \sim 1\,deg\,s^{-1}$.
After our association of XSS\,J1227 to the \emph{Fermi} source (dM10), we were granted \emph{AGILE}/GRID
data in Cycle\,3 (Prog.ID:70). We have enlarged the dataset including observations
collected in pointing mode from Oct.\,1, 2007 to Oct.\,31, 2009, and the data collected with
the satellite in spinning mode from Oct.\,31, 2009 to May\,15, 2011; the total exposure times amount 
to 5.3\,Ms and 3.8\,Ms respectively.

\begin{figure}[h!]
\resizebox{\hsize}{!}{\includegraphics{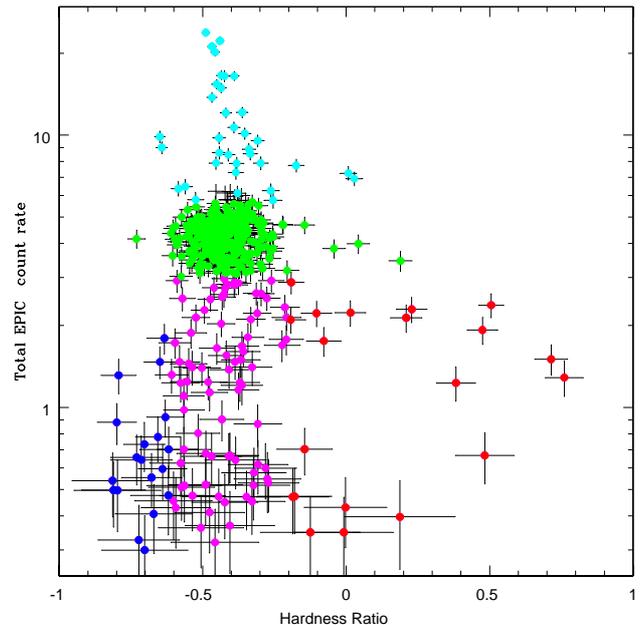}}
\caption{HID  between 0.3-2\,keV and 2--10\,keV EPIC 
bands.  Flares are marked in cyan, dips with no spectral change in  magenta, soft  dips in blue,
 and hard dips  in red. Green 
points  represent quiescence. See electronic version for colour figure. }
\label{HR_PN}
\end{figure}

XSS\,J1227 is too faint in the X-rays to be detected by the Super-AGILE instrument  and therefore only the GRID dataset
was analysed using the Build\,20  release  of the GRID analysis software available at the ASI--Science Data Centre (SDC). 
Well-reconstructed $\gamma$-ray events  were selected using the FM3.119 filter. Events collected during the passage 
in the 
South Atlantic Anomaly and Earth-albedo gamma rays coming from a circular region of
radius $\rm 80^{\circ}$ centred on the Earth were rejected. A maximum likelihood analysis was 
performed in the 
100\,MeV--50\,GeV energy band, obtaining  2$\sigma$ upper limits of 
$\rm 7\times 10^{-8}$\,photons\,cm$^{-2}$\,s$^{-1}$ with a total exposure coverage of $\rm 2\times 10^9$\,cm$^2$\,s
for the pointing-mode period and  $\rm 1.2\times 10^{-7}$\,photons\,cm$^{-2}$\,s$^{-1}$ with a total 
exposure 
coverage of 
$\rm 1.5\times 10^9$\,cm$^2$\,s for the spinning-mode period. For the total period under analysis we 
obtained a 
2$\sigma$ upper limit of  $6\times  10^{-8}$\,photons\,cm$^{-2}$\,s$^{-1}$ with a 
total exposure coverage of $\sim 3.4\cdot 10^9$\,cm$^2$\,s. In order to investigate soft (E$<$200MeV)
gamma-ray emission from XSS\,J1227, we also integrated the  GRID data in the non-standard 50--200\,MeV 
energy band, 
obtaining a  2$\sigma$ upper limit of $14.8\times 10^{-8}$\,photons\,cm$^{-2}$\,s$^{-1}$
with a total exposure coverage of $2\times 10^9$\,cm$^2$\,s.

\subsection{The \emph{Fermi} data}

The \emph{Fermi}-LAT data  were retrieved from the \emph{Fermi} Science Support Centre.
The dataset spans 44  months since the start of \emph{Fermi} operations, i.e. from Aug.4,\,2008 to 
Apr.17,\,2012, 
 amounting  to a total exposure time of 52.6\,Ms.
We used Pass7 photon data, reducing and analysing them using the Fermi Science Tools v9r27 package. 
We used the high-quality 
(diffuse) photon event class (EVENT CLASS = 2) and the Pass 7 v6 Source (P7 V6 source) instrument 
response functions (IRFs). We excluded time periods when the region around 2FGL\,J1227 was 
observed  at a zenith angle greater than 105$^o$ to reduce contamination by Earth albedo gamma rays. 
Correction to the solar system baricentre was also applied.
 
We produced a LAT aperture photometry light curve with a time bin of four days from a circular region 
centred 
on the source position and a radius of $\rm 1^{\circ}$ to avoid contamination from close sources (2FGL\,J1218.8-4827, 
2FGL\,J1231.3-5112 and 2FGL\,J1207.3-5055) (see \citet{Nolan12}). We performed  this operation with 
the \textit{gtbin}  task and applied the exposure correction using the \textit{gtexposure} task. 
These tasks do not perform background subtraction, and hence no background correction  was applied.

\begin{figure*}[th!]
\resizebox{\hsize}{!}{\includegraphics{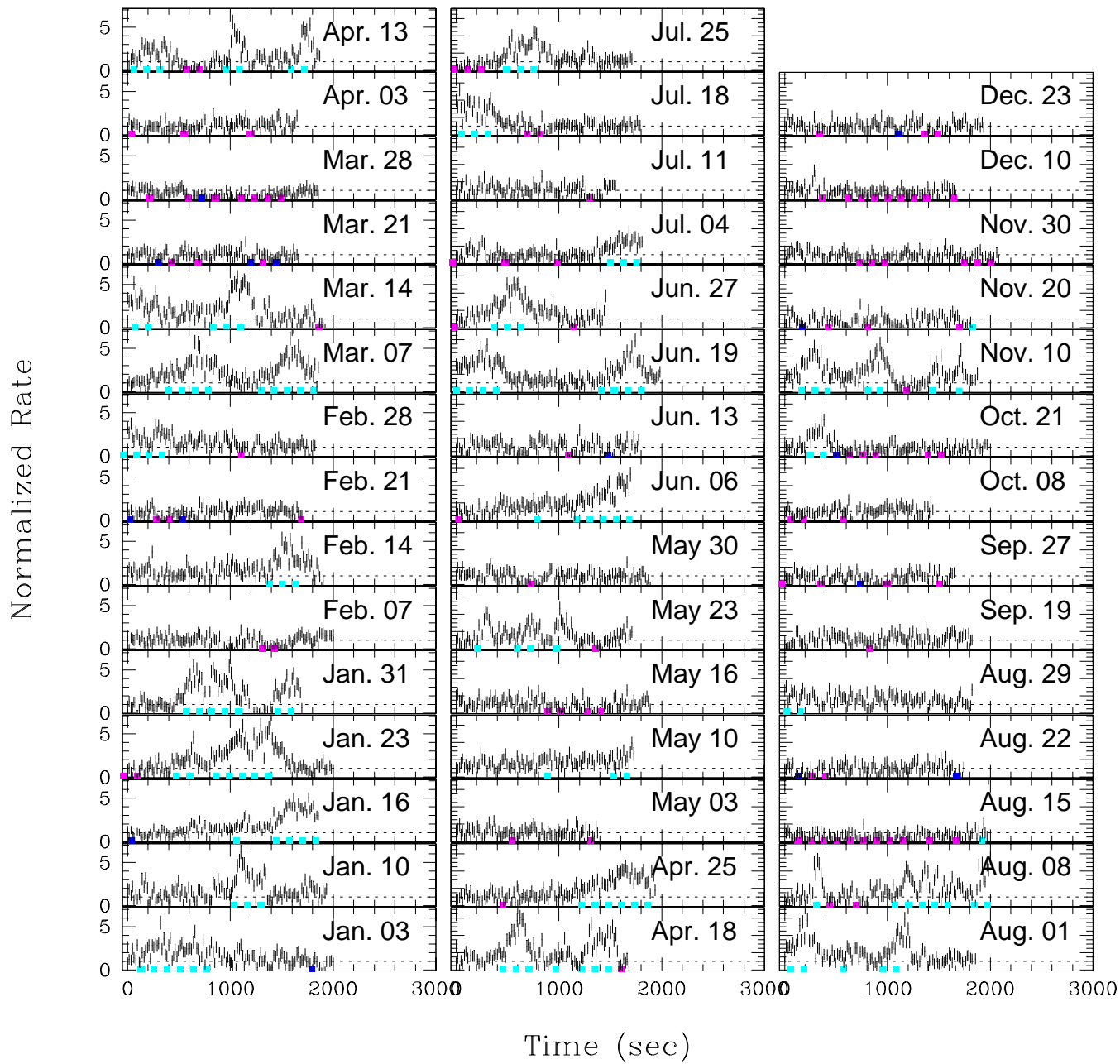}}
\caption{ \emph{RXTE} net light curves during the pointings in 2011 with a binning time of 
16\,s. The count rate is normalized to the quiescent average value observed on Dec.\,23. Flares and 
dips  are shown in colours as in Fig.\,1. See electronic version for colour figure.} 
\label{RXTElc}
\end{figure*}

The spectral fluxes in the five LAT bands (100-300\,MeV, 0.3--1\,GeV, 1--3\,GeV, 3--10\,GeV, and 
 10--100\,GeV) were instead collected from the $\rm 2^{nd}$ source catalogue that covers 
24\,months of operations and were obtained using maximum likelihood analysis (see details in 
\citet{Nolan12}). 2FGL\,J1227 is reported   at a flux of 
3.34$\pm$0.23$\times  10^{-11}$\,erg\,cm$^{-2}$\,s$^{-1}$ in the 0.1--100\,GeV range at a significance of 24.3\,$\sigma$
in accordance with the value of 3.95$\pm$0.44$\times  10^{-11}$\,erg\,cm$^{-2}$\,s$^{-1}$ reported in the $\rm 1^{st}$ 
LAT source catalogue. 
The  localization  procedure used to construct the catalogue also provides spectral fits to all sources. 
In contrast to the $\rm 1^{st}$  catalogue, which used only a power-law function, 
different spectral shapes are used in the $\rm 2^{nd}$ one. For 2FGL\,J1227,  a 
power-law and a 
function, called ''logParabola" are fitted to the spectrum. This 
representation, allows for a smoother decrease at high energy than a power-law exponential cutoff 
form. It reduces to a 
simple power-law when the curvature parameter $\beta$=0 and is reported in the catalogue when the curvature
is significant by above $4\sigma$ (\citet{Nolan12}). A simple power-law fit is reported with a 
spectral index of 2.3306. The spectral index of the 
logParabola is 2.085$\pm$0.087, $\beta$=0.288$\pm$0.064 
with a pivot energy at  549.30\,MeV (where the uncertainty on differential flux is minimal). The 
significance of the 
fit improvement with respect to the simple power-law is 5.6$\sigma$.

\begin{figure}[h!]
\resizebox{\hsize}{!}{\includegraphics{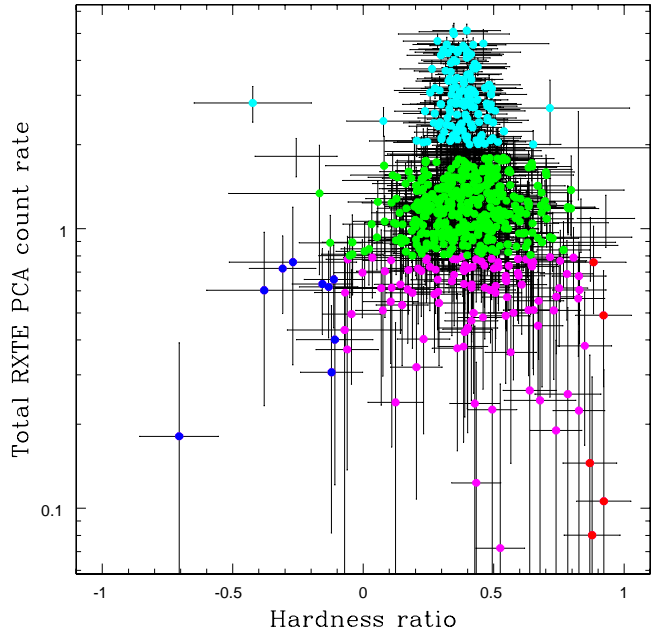}}
\caption{ HID between the  $\rm HR_{RXTE}$ 2--4\,keV and 4--9\,keV bands as defined in the text. 
Colours are used as in Fig.\,2. See electronic version for colour figure. }
\label{RXTE_HR}
\end{figure}

\section{Results}

\subsection{X-ray persistent characteristics}

The background-subtracted X-ray light curves from EPIC-pn  and MOS  cameras were first extracted in 
60\,s bins to evaluate the  overall behaviour of XSS\,J1227 during the \emph{XMM-Newton} observation. The light 
curves from the EPIC-pn and MOS are similar, and we use the EPIC-pn data because of the higher S/N.
In Fig.~\ref{HR_time_PN}, the 0.3--10\,keV light curve with a binning time of  
10\,s is shown together with the temporal behaviour of the hardness ratio HR = [H-S/H+S], where S and 
H are the  count rates in the 
0.3--2\,keV and 2--10\,keV, respectively, with a binning time of 60\,s.  XSS\,J1227 was found to be 
at a 
persistent (quiescent) level of $\rm \sim4.35\pm 0.35\,cts\,s^{-1}$ for most of the 
EPIC-pn exposure,  except for the last 4\,ks, where the count rate increases up to a factor of 6.4. 
While the general behaviour observed in 2011 is similar to 
that of the  Jan. 2009 \emph{XMM-Newton} observation (dM10), the EPIC-pn timing mode data show that the 
persistent level is characterized by  dips of  variable length (see Sect.\,3.3).

The hardness ratios confirm  the general trend observed in 2009 (dM10), where a  substantial 
hardening is 
detected  during the  postflare dips and there is no major spectral change during flares and 
quiescent dips.  
The intensity-versus-hardness ratio diagram (HID) in Fig.~\ref{HR_PN} is similar (see Fig.\,4 of 
dM10).
Here we define flares when the count rate $>$5.7\,cts\,s$^{-1}$ (cyan in Fig.~\ref{HR_PN}) and dips
have a count rate $<$3.\,cts\,s$^{-1}$. The higher S/N EPIC-pn data allow the dips to be better 
sampled. Therefore we further
separate soft  dips with HR$<$-0.6 (blue) and hard dips with HR$>$-0.2 (red) from ''ordinary" 
neutral dips with -0.6$<$HR$<$-0.2 (magenta). The hard and soft dips are defined as those below/above 
2$\sigma$ the average HR value in 
quiescence. The HID confirms  that only post-flare dips are hard, while dips during quiescence are 
mostly neutral with a marginal softening in those that are longer.

Flares and dips in the  X-ray light curve of XSS\,J1227 are found in all observations conducted in 2011 by 
\emph{RXTE} and \emph{Swift}. In  Fig.~\ref{RXTElc}, we show the light curves of each 
\emph{RXTE} pointing with a binning time of 16\,s. The hardness ratios  $\rm 
HR_{RXTE}$=[H-M/H+M] in  the 2--4\,keV and the 4--9\,keV bands were constructed with  a binning time of 128\,s.
We  define the 
quiescent level as that observed on Dec.\,23  (Table\,\ref{observ}), and hence flares are defined 
 with a count rate $>$ 1.8\,cts\,s$^{-1}$ and dips 
with  a count rate $<$0.8\,cts\,s$^{-1}$. The  average hardness  ratio in quiescence is $\rm 
HR_{RXTE}=0.38\pm$0.24, and 
we define  hard/soft dips as those over/below 2$\sigma$ this average value, 
respectively. Ordinary neutral dips are instead those found within this range. The \emph{RXTE} HID 
(Fig.~\ref{RXTE_HR}) shows that flares and ordinary and soft dips have been relatively well sampled, 
but not
the hard dips. This is likely due to the higher energy coverage of \emph{RXTE}/PCA with respect to \emph{XMM-Newton} and
\emph{Swift} instruments. 

\begin{figure}
\resizebox{\hsize}{!}{\includegraphics{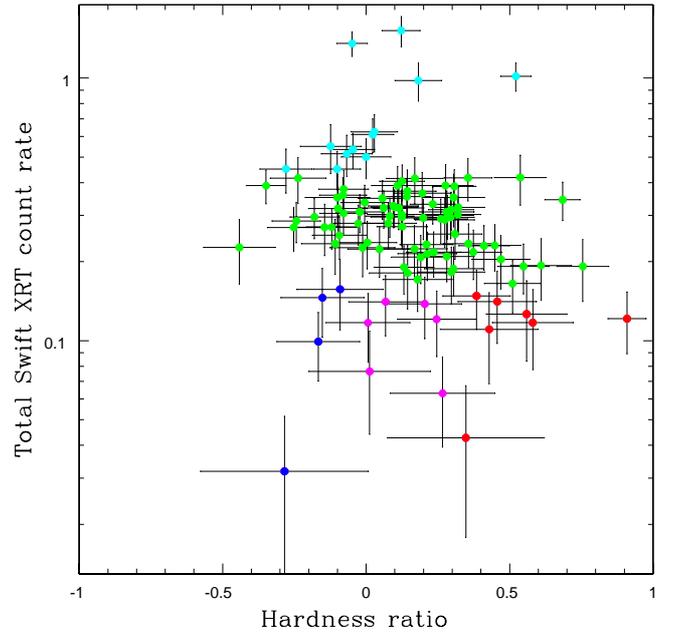}}
\caption{HID between  the 0.3--1.5\,keV and 1.5--10\,keV \emph{Swift}/XRT bands. 
Colours are used as in previous figures. See electronic version for colour figure. }
\label{Swiftlc}
\end{figure}

The \emph{Swift} light curves (not shown) were extracted with a binning time of 25\,s. 
Again, XSS\,J1227 is found stable at $\sim$0.2-0.3\,cts\,s$^{-1}$ during most of \emph{Swift} pointings
(see Table\,\ref{observ}).
Despite the poor  coverage,  several flares are detected or partially detected, with 
one notably on Sept.\,19 that reached peak intensity $\sim$8 times the quiescent 
(0.26\,cts\,s$^{-1}$) level. 
The flares are defined with a count rate 
$>$0.43\,cts\,s$^{-1}$, while dips are defined 
with a count rate $<$0.2\,cts\,s$^{-1}$. The hardness ratios 
were constructed with a binning time of 128\,s 
in the 0.2--1.5\,keV and 1.5--10\,keV bands. Soft/hard dips were defined as over/below 2$\sigma$ the 
average  value $\rm HR_{Swift}$ = 0.1$\pm$ 0.1. The HID from \emph{Swift} also reveals the tendency 
of  hardening during post-flare  dips (Fig.~\ref{Swiftlc}). 

We also inspected archival \emph{Swift} unpublished observations on Sept.\,15, 2005, Sept.\,24, 2005 
and Aug.\,10, 2010  in similar  fashion as the 2011 \emph{Swift} pointings to construct a long--term 
history of XSS\,J1227.

The data acquired during the 2011 monitoring compared with the observations carried out in 2005 by 
\emph{Swift}, in 2007 by \emph{RXTE}  (dM10), in 2008 by \emph{Suzaku} 
\citep{Saitou}, in 2009 with \emph{XMM-Newton} (dM10) and  \emph{RXTE} 
\citep{Saitou11}, and in 2010 by \emph{Swift} allow to definitively assess that XSS\,J1227 
is a persistent and not a transient source.
 It is stable on a 7--yr timescale 
while undergoing repetitive flaring episodes. 
From the longer datasets we also  estimate flare occurrence and duration: $\rm \Delta_{Quie} \ga$24\,ks and 
$\rm \Delta_{Flares} \ga$ 8.2\,ks from \emph{Suzaku} (2008); $\rm \Delta_{Quie} \ga$35\,ks and $\rm \Delta_{Flares} \ga$ 
3.6\,ks  from \emph{RXTE} (2007) and $\rm \Delta_{Quie}=$27.1\,ks and $\rm \Delta_{Flares} \ga$ 3.1\,ks from 
\emph{XMM-Newton} (2009) and $\rm \Delta_{Quie}\ga$27.4\,ks and $\rm \Delta_{Flares} \ga$ 3.4\,ks from \emph{XMM-Newton}
(2011), giving lower limits of $\rm \Delta_{Quie}=$27-35\,ks  and $\rm \Delta_{Flares} \sim$ 4\,ks, the latter 
grouped in multi-flares. These could be a signature of a  $\ga$9-11\,h occurrence.

\subsection{The near-UV light}
Figure\,\ref{HR_time_PN} also reports the simultaneous light curve 
recorded with the OM in the U band 
filter in Jan.\,2011,   where the source also undergoes a flaring activity 
at the end of the \emph{XMM-Newton} 
observation.
XSS\,J1227 is at an average U band magnitude of 16.6\,mag during quiescence and reaches 15.1\,mag at flares. The highest 
peak in the U light curve is $\sim$3 times the persistent level,  smaller than the X-ray variations.
Despite the gaps in this light curve, it is 
apparent that dips occur also in the near-UV during quiescence,  with the count rate dropping by a factor of
$\sim$1.4. Hence, in contrast to  what was claimed in dM10, it is now possible to identify the UV 
counterparts  of the X-ray dips in  XSS\,J1227.  An enlargement of the light curves is shown in Fig.\,\ref{UV_X_enlarg_norm}.

Furthermore, a long-term variation in the UV was suspected  to be present in the 2009 
\emph{XMM-Newton} OM data.
A Scargle periodogram of the quiescent U band time series shows a peak at 6.7\,h (at 3.6$\sigma$ c.l.). 
A formal sinusoidal fit gives a period of 6.4$\pm$0.2\,h (uncertainty at 1$\sigma$ c.l.) and full amplitude of 22$\pm$1$\%$ 
(see  also Fig.~\ref{HR_time_PN}). The statistical significance of the inclusion of a sinusoid with respect to a 
constant function,  evaluated with a F-test, is 10$\sigma$. Given that the length of the U band exposure during quiescence
 is  26\,ks, the 6.4\,h period should be regarded as a lower limit.
We note here that, in contrast to what was claimed from the past \emph{XMM-Newton} pointing (dM10),  
no variations on timescales of hours are found in this EPIC-pn quiescent light curve.

\subsection{The dips in focus}

 The X-ray dips represent a persistent characteristic of XSS\,J1227 during quiescence. The time spent in the 
dips amounts to   7.78\,ks, representing $\sim$30$\%$ of quiescent period.  A few 
(ten are marked in 
Fig.\,\ref{HR_time_PN}) longer dips have durations  between 100\,s and 700\,s, but the one observed 
prior the flaring 
activity is the longest, with a duration of 1100\,s. Similarly, in the 2009 \emph{XMM-Newton} 
pointing, the longest dip is 
detected before the flaring activity. In both observations, these dips are soft. 
The source also  undergoes several short dips of length 10-90\,s, where  the 
count  rate does not drop to zero.   These short dips are the ordinary neutral dips with no substantial spectral change 
(Fig.\,\ref{HR_PN}).

\begin{figure}
\resizebox{\hsize}{!}{\includegraphics{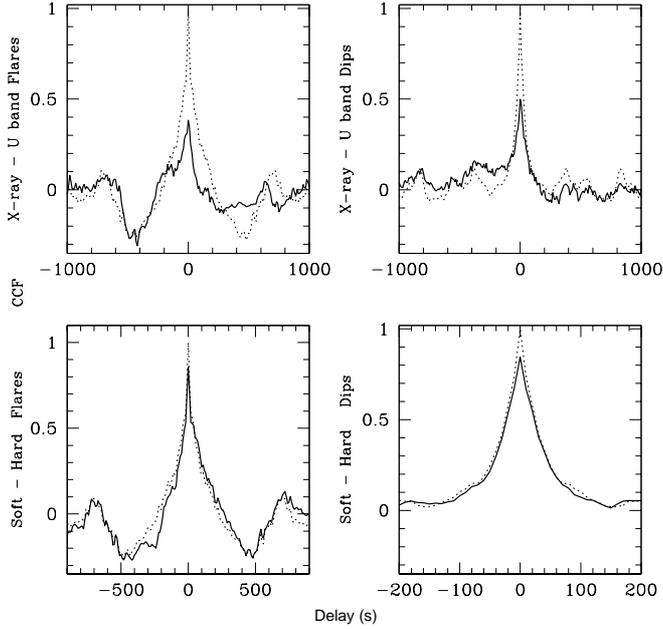}}
\caption{ CCFs (solid lines) during flares (left) and dips (right). The CCF between the hard and the soft X-ray  
bands, taking as reference the soft range (ACF), are shown in the bottom panels.  The  CCFs between 
the total X-ray 
and U bands, taking as reference the X-ray band (ACF), are shown in the top panels. 
The ACF of each reference band  is plotted with a dotted line. }
\label{cross_corr_2011_flares_dips}
\end{figure}

The selected ten long dips are characterised 
 by a less steep, though rapid,  decay reaching almost 
 zero  counts in the longer ones and a faster recovery to the persistent level. The decay and rise  times 
have  lengths ranging between $\sim$40-90\,s and appear to be correlated with the length of the dips, with
slower decays/rises in the longer 
dips.  The hardness ratios show a softening only in the longer dips. 
We also performed a cross correlation (CCF) between the  hard  (2--10\,keV) and  soft  (0.3-2\,keV) 
light curves during the ten selected long dips. The CCF  
is shown in the right-hand lower panel of Fig.\,\ref{cross_corr_2011_flares_dips}, where the soft 
band is taken as a reference and
its autocorrelation function (ACF) is also reported. The CCF   does not  reveal any lag 
 between the two bands, implying that the bulk of X-ray flux arises from the same region.

\begin{figure}
\resizebox{\hsize}{!}{\includegraphics{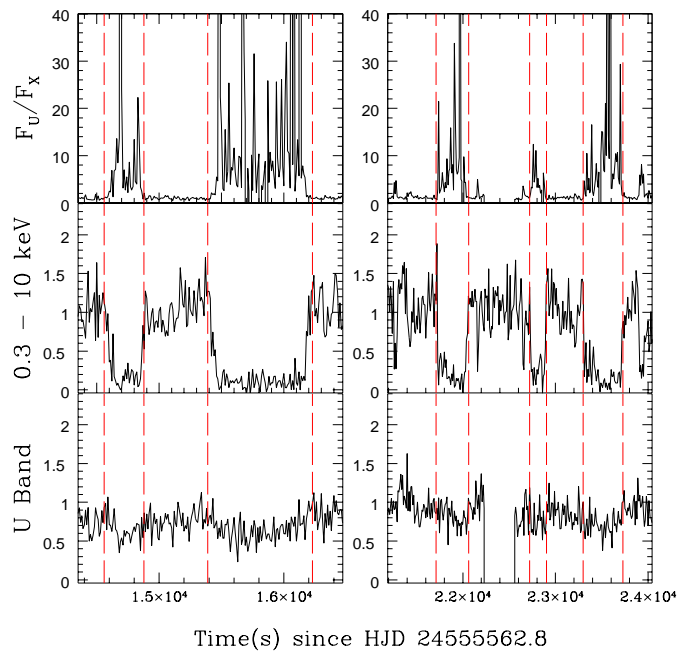}}
 \caption{Enlargement of the normalized light curves in 2011 showing the simultaneous occurrence of quiescent dips in 
the U band  (bottom panel) and X-rays (middle panel) (0.3-10\,keV)  with a
binning time of 10\,s, together with the ratio of the U and 
X-ray band count rates (top). Vertical dashed (red) lines mark the dip 
start and 
end times.}
\label{UV_X_enlarg_norm}
\end{figure}

Near-UV  dips seen in quiescence occur almost simultaneously with the X-ray dips, but the flux deficiency is 
$\sim40\%$, hence
much lower than their X-ray counterparts  (Fig.\,\ref{UV_X_enlarg_norm}). The shape
of UV dips is  also different, being shallower with a smooth decay and rise. This is reflected  in 
the count rate ratios between
the two  bands. Using the total X-ray range as reference, we computed its ACF and cross-correlated 
the U band light curve with the X-ray one using eight of the ten dips covered 
simultaneously  (see Fig.\,\ref{HR_time_PN}). The CCF is shown in the right-hand top panel 
of Fig.\,\ref{cross_corr_2011_flares_dips} together with the ACF of the X-ray band. It is peaked at zero 
lags, but an excess at negative lags $\sim$100-300\,s is found. Both the UV deficiency and CCFs 
indicate that the UV-emitting region is more extended.

\begin{figure}
\resizebox{\hsize}{!}{\includegraphics{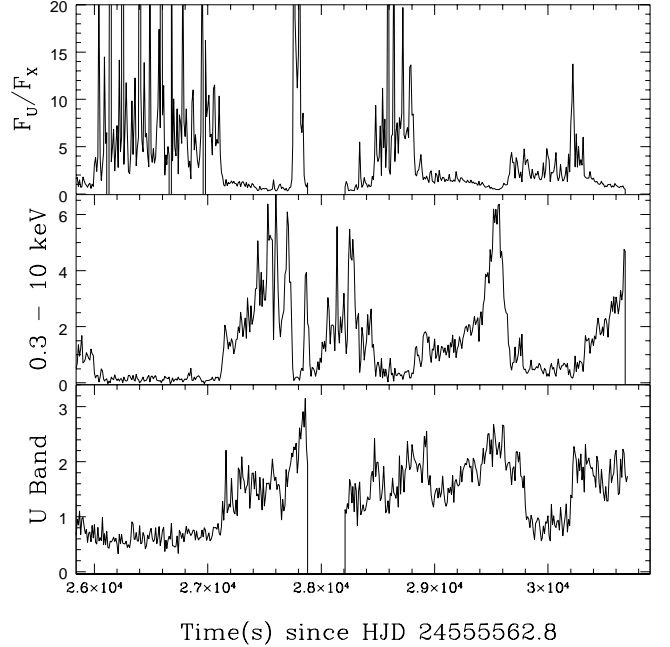}}
 \caption{ Enlargement of the normalized light curves showing the flaring activity  in the U band 
(bottom panel) and X-rays (middle panel) (0.3-10\,keV)  with a
binning time of 10\,s, together with the ratio of the U and X-ray band (top).}
\label{UV_X_flares_norm}
\end{figure}

\subsection{The flaring states}

 XSS\,J1227 showed flares at the end of the  \emph{XMM-Newton} observation where the X-ray count 
 rate increased by a factor of $\sim$6.4 (Fig.\,\ref{UV_X_flares_norm}). Similar
 to the 2009 \emph{XMM-Newton} light curve,  the flaring 
activity  consists of multiple events. Here  we identify at least four major events, lasting about 
600\,s, 640\,s, 847\,s, and 
$>$350\,s (the latter covered only  at the start). The higher S/N EPIC-pn timing mode 
data, while confirming  the general exponential rise and rapid decay of major flares, also allow us
to  identify 
that they are highly structured and composed of multiple peaks with a typical timescale of 30-50\,s. 
The major 
spectral changes occur during the dips after these flares, as depicted in Fig.\,\ref{HR_time_PN}, 
where  
hardening is detected but not at flare rise and maximum. 
We cross-correlated the hard (2-10\,keV) with the  soft (0.3-2\,keV, taken as reference) bands 
and computed the ACF  in the soft band in the flare portions during the rise and 
maximum. The CCF and ACF are
shown in Fig.\,\ref{cross_corr_2011_flares_dips}, where no delay is inferred,  confirming what 
was found in 2009.

The flares, simultaneously detected in the U band, display peak intensities up to a factor of $\sim$3.
In the U band, only one post-flare dip can be recognized precisely after the third flare 
(Fig.\,\ref{UV_X_flares_norm}).
The CCF between U and X-ray (taken as reference)  peaks at zero lags but is also 
asymmetric towards negative lags ($\sim$150-200\,s). It is different
from what was found in 2009, possibly because of the short 
coverage between the two bands  in that observation. This could be due to the rather different structure of U flares
with respect to the X-ray ones,   where not all post-flare
X-ray dips have their UV counterpart and the last UV flare rises faster than the X-ray one.

\subsection{Search for fast coherent signals}

\begin{figure}
\resizebox{\hsize}{!}{\includegraphics{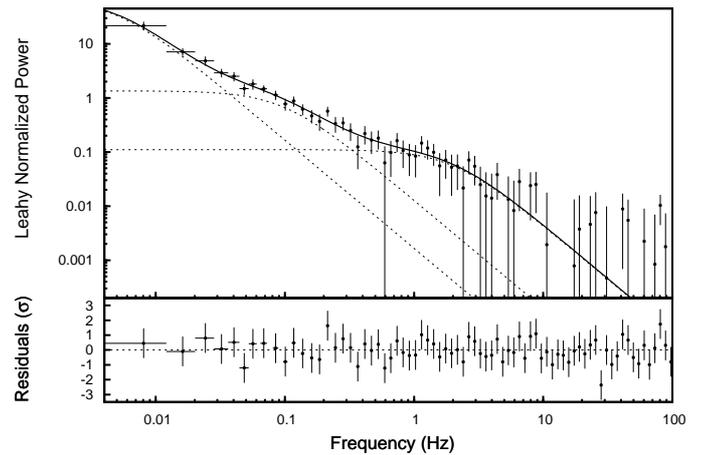}}
\caption{{\it Top panel:} Power spectrum obtained by averaging 242 intervals, each
  $\simeq 124$\,s long, and logarithmically re-binning the 
  spectrum with a factor of 1.1. The white noise level
  of $1.9880(4)$, evaluated by modelling  the spectrum between 0.5  
  and 2 kHz with a constant, is subtracted from each power estimate. 
{\it Bottom panel:} Residuals in units of $\sigma$ of the power  spectrum with
  respect to the best-fitting model composed of three lorentzians
  centred at zero frequency.}
\label{avg_psd}
\end{figure}

The EPIC-pn timing mode data were also inspected to search for fast periodic or quasi-periodic 
signals in the range 0.5-10\.keV. 
In order to assess the presence of a coherent signal in the EPIC-pn
observation of {\xss}, a Fourier analysis was first performed using the
whole exposure, binning the time series at eight times the temporal resolution of the EPIC-pn timing 
mode data, 
so that $\rm \mbox{t}_{bin}\simeq0.236$\,ms. Hence the search for periodic
signals was performed in the frequency interval $3.3\times10^{-5}$ -- $2114$\,Hz. 
We took into account the number of frequency bins searched to define a
threshold power level to be exceeded in the presence of a signal above
the noise \citep[see, e.g.][]{vdK89}. Because of the presence of red
noise at low frequencies (Fig.~\ref{avg_psd}), we used a    
smoothing window technique to evaluate the detection level at each of
the frequencies searched \citep{IsrStl96}. No signal was significantly
detected at a confidence level of 3$\sigma$. Given the maximum power
detected in the power spectrum, we evaluated an upper limit at 3$\sigma$  
confidence level on the fractional amplitude of pulsations of
$\rm A_{UL}^{is}=0.050$. The effects of the observation length and
binning, as well as the interaction between a putative signal and
noise, are accounted for as described by \citet[]{Vgh94}. The power spectrum continuum below 100 Hz is 
shown in Fig.\,\ref{avg_psd}. Three flat-top noise components of width,
$W_1=(1.0_{-0.5}^{+0.7})\times10^{-2}$, $W_2=0.19_{-0.05}^{+0.08}$ , 
and $W_3=4.0_{-1.0}^{+1.5}$ Hz, adequately fit the power
spectrum ($\chi^2=37/67$).

The upper limit on the pulsation amplitude is meaningful
only if a putative pulsar is isolated or belongs to a wide binary
system ($\rm P_{orb}>>T_{obs}$). In dM10 we reported a possible 4.32$\pm$0.01\,h 
optical photometric periodicity, possibly associated with the binary orbital period. 
The \emph{XMM-Newton} OM U band light curve indicates the presence of a longer period
6.4$\pm$0.2\,h (see also Sect.\,4). If the binary period is as short as  4.3-6.4\,h,
the secondary star mass would be in the range 0.3-0.7\,M$_{\odot}$ \citep{Smith_Dhillon98}.
We therefore need to consider shorter integration times  in order to limit the leakage of power in close
frequency bins due to the Doppler shift affecting the signal
frequency.  The optimal integration time to  search for a signal coming from a binary system, without making any
correction of the photon arrival times for the unknown orbital
motion, was evaluated by \citet{JhnKlk91}. 
 Adopting as a lower limit the 4.3\,h period and using the corresponding  orbital parameters, one obtains

\begin{center}

\begin{equation}
\rm  T_{best}\simeq 247 \left(\frac{\nu}{300 \mbox{Hz}}\right)^{-1/2} \left(\frac{P_{\mbox{orb}}}{4.3 
\mbox{h}}\right)^{2/3}
\left(\frac{M_2}{0.3 \mbox{M}_{\odot}}\right)^{-2/5} \left(\frac{\sin{i}}{\sin{45^{\circ}}}\right)^{-1/2}\mbox{s}.
\end{equation}
\end{center}

Assuming that a pulsar in this LMXB system has a frequency in the
range displayed by the known accreting pulsars with such low mass
companions (100--600 Hz), one obtains optimal integration times
spanning 150--395\,s. We produced several power spectra over time intervals of
length 124 and 248\,s~\footnote{The lengths of the time intervals 124\,s, 248\,s, and 1983.7\,s were 
chosen in 
order to give a number of 
bins equal to an integer power of 2$^{19}$, 2$^{20}$, and 2$^{23}$, respectively. This greatly 
decreases the CPU time needed to perform fast Fourier transforms (FFT)}. 
No significant detection within a 3\,$\sigma$
confidence level was obtained. However, this 
is not surprising however, because at the low count rates of this 
source, the number of counts in every interval is so low that only extremely loose  
3$\sigma$ upper limits on the pulse amplitude of the order of $\rm A_{UL}\sim0.5$
could be obtained.

\begin{figure}
\resizebox{\hsize}{!}{\includegraphics{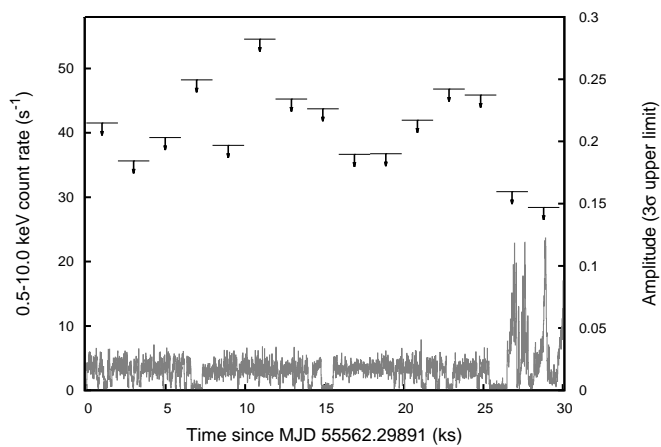}}
\caption{Right scale: upper limits at a 3$\sigma$ confidence level  
  on the pulse amplitude  during the EPIC-pn observation of
  {\xss}, as obtained by performing a QCRT over 1983.7 s long
  intervals (see text). Left scale: the net 0.5--10 keV light curve recorded by the EPIC-pn is also 
over-plotted.}
\label{UL}
\end{figure}

In order to draw more stringent constraints on the presence of a
coherent signal in the dataset, we applied quadratic coherence
recovery techniques \citep[QCRT;][]{Wod91,Vgh94}. Such techniques rely
on the correction of the times of arrival of the X-ray photons by     
using a quadratic time transformation, under the assumption that the
sinusoidal Doppler modulation of the arrival times introduced by the
orbital motion is well approximated by a parabola over the considered
integration time. The optimal integration time for the detection of a
signal with QCRT was derived by \citet{JhnKlk91} as

\begin{center}
\begin{equation}
\rm  T_{best}^{acc}\simeq 1361 \left(\frac{\nu}{300 \mbox{Hz}}\right)^{-1/3} \left(\frac{P_{\mbox{orb}}}{4.3 
\mbox{h}}\right)^{7/9}
\left(\frac{M_2}{0.3 \mbox{M}_{\odot}}\right)^{-4/15} \left(\frac{\sin{i}}{\sin{45^{\circ}}}\right)^{-1/3}\mbox{s}.
\end{equation}
\end{center}

We therefore split the data in $N_{ \mbox{intv} } = 15$ intervals composed of $2^{23}$ bins, 
each of length equal to $\mbox{t}_{bin}$ as before, and performed   
FFT on the time series corrected with the relation
$\rm t'=\alpha t^2$, where $\rm \alpha=-1/2 \:(a \sin{i}/c)\: (2\pi \nu)
\:(2\pi/P_{orb})^2\: \sin{\phi_0}$, $\rm a \sin{i}/c$ is the projected
semi-major axis of the primary star orbit and $\rm \phi_0$ is the orbital
phase at the beginning of the time series. The maximum and minimum
values of $\alpha$, as well as the value $\delta\alpha$ by which
$\alpha$ is incremented in each of the corrections, have been
determined by \citet{Vgh94}  as $\rm \alpha_{max}=-\alpha_{min}=8.80\times10^{-8}    
(q/0.29) (P_{orb}/4.3hr)^{-4/3} (M_{tot}/1.7M_{\odot})^{1/3}\:s^{-1}$,
and $\delta\alpha=1/(2\nu_{Ny} T^2)$. By considering the range
of values for the parameters of the binary system  and  the   
reported uncertainties, a range of masses between 1.4 and 2.0   
M$_{\odot}$ for the mass of a putative NS in this system,
$\nu_{Ny}\simeq2114.35$ Hz and $\rm T_{FFT}\simeq1983.7$ s, we set
$\rm \alpha_{max}=-\alpha_{min}=1\times10^{-7}$ s$^{-1}$ and
$\delta\alpha=6\times10^{-11}$ s$^{-1}$ so that $N_{\alpha}=3334$
corrections were performed on each of the $N_{intv}=15$ time
intervals. The detection threshold is therefore $P_{det}=62.58$ in    
the frequency range, where the noise is compatible with being
distributed as a chi-squared with two d.o.f. (see \citet{Vgh94}). No signal   
above this level was found. The upper limits at 3$\sigma$ confidence
level set on the pulse amplitude in each of the intervals are plotted
in Fig.~\ref{UL}, together with the  light curve in the 0.5--10 keV band.
The values we find as limits on the pulse amplitude ($\sim 0.15-0.25$)
are larger than those usually measured from pulsars in
LMXB systems.

\subsection{Persistent X-ray and gamma-ray emissions}

The X-ray history of XSS\,J1227 collected over 7\,yr shows that it is a persistent and rather 
stable X-ray source, with  flares and dips being peculiar and permanent characteristics. 
 The associated  \emph{Fermi}-LAT source 2FGL\,J1227 is reported in both $\rm 1^{st}$  and $\rm 
2^{nd}$ catalogues
to be non-variable  on a  monthly  timescale over the first two years of \emph{Fermi} operations  (see \citet{Nolan12}) and 
also  
on timescale $>$ 4\,d in the
0.1--300\,GeV range over the first $\sim$25.5\,months of \emph{Fermi}-LAT  observations 
\citep{Hill}.

\begin{figure}
\includegraphics[width=\columnwidth,height=7.cm]{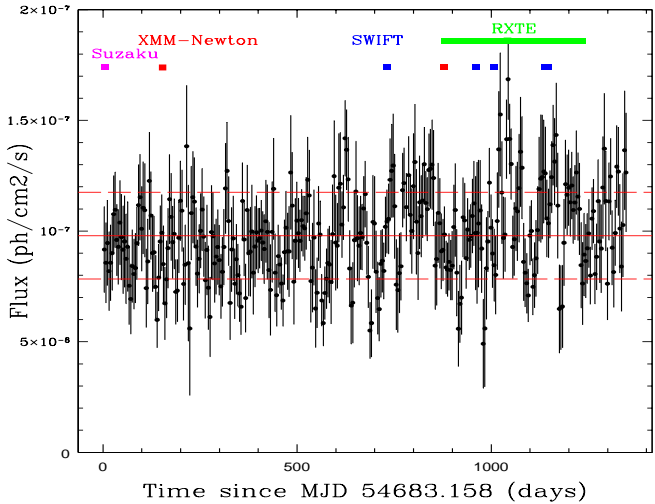}
\caption{ \emph{Fermi}-LAT light curve in the range 0.1--300GeV with a 4\,d binning time from 
Aug.2008 
to Apr. 
2012 extracted in a 1$^o$ aperture. The mean flux is reported together with 
the 1$\sigma$ standard deviation. The background is not subtracted. The X-ray coverages by \emph{Suzaku} (magenta), \emph{RXTE} 
(green),  \emph{XMM-Newton} (red),  and \emph{Swift} (blue) are also reported. See electronic version 
for colour figure.}
\label{LAT_X_curve}
\end{figure}

We then extended the gamma-ray coverage in the 0.1--300\,GeV range up to Apr.\,18, 2012 and show the 
whole 
light curve (Fig.\,\ref{LAT_X_curve}) with the same 4\,d binning as \citet{Hill}.
The average photon flux is $\rm \sim 1\times 10^{-7}\,photons\,cm^{-2}\,s^{-1}$, 
slightly lower but consistent with that obtained by \citet{Hill}. Differences are due to the different
 photon selection criteria and IRF adopted at the two epochs of data analysis~\footnote{Details of reccomendation for data 
selection  at the Fermi Science Support Centre: 
http://fermi.gsfc.nasa.gov/ssc/data/analysis/documentation/Cicerone\, /Cicerone$\_$Data$\_$Exploration/Data$\_$preparation.html}
We have also overlaid  
the X-ray pointings since 2008, from
which it is possible to conclude that  both  X-ray and gamma-ray emissions are detected simultaneously.
Since a shorter time bin size is precluded by the low photon flux, it is not possible to infer 
 whether the gamma rays are emitted steadily or bursts or flares.

\section{The spectral characteristics}

In dM10 we reported the analysis of the average X-ray spectrum in the 0.3--100\,keV range using \emph{XMM-Newton}, 
\emph{RXTE}, and \emph{INTEGRAL}. The spectrum is featureless and well fitted with an absorbed 
power-law with 
$\Gamma$=1.7$\pm$0.02. Similar results were found by \citet{Saitou}. The spectral shape was also found to be 
invariant to flux variations, e.g. during quiescence, quiescent dips, and flares, but not during the 
hard post-flare  dips, where a partial covering absorber better described the spectrum (dM10). We 
have further checked this behaviour 
using the 2011 
\emph{XMM-Newton} EPIC-pn data, fitting them with an absorbed power-law and obtained similar results: 
$\Gamma$=1.64$\pm$0.01 (quiescence), $\Gamma$=1.65$\pm$0.03 
(flares), $\Gamma$=1.71$\pm$0.04 (quiescent dips), and $\Gamma$= 
0.74$\pm$0.08 (post-flare dips). The  column  density of the absorber is within errors the same in all fits 
and  similar to that found in the 2009 
 \emph{XMM-Newton} data, $\rm N_H = 0.75\pm0.04 \times 10^{21}\,cm^{-2}$, and consistent with the 
galactic 
medium in the direction of the source. The fit to the post-flare dip spectrum improves when adding a 
partial covering absorber. We find $\rm \chi^2/d.o.f.$=33/25 vs 
$\rm \chi^2/d.o.f.$=43/27 with and without this component. This component is found to be significant at the 
2.1$\sigma$ confidence level. 
The column density and partial 
covering fraction of the absorber are $\rm 3.85^{+2.5}_{-1.4} \times 10^{21}\,cm^{-2}$ and 
0.61$\pm$0.2, respectively. 
Also,  fits to the  \emph{RXTE} quiescent and flare spectra give similar  power-law indexes,  
further corroborating the unvariancy of the spectrum during active and persistent level. 
The spectrum of the quiescent dips might indicate a  softening, but it does not require an 
additional soft component.The 2--10\,keV fluxes as derived for the 2011 \emph{XMM-Newton} data  
are 0.26$\pm$0.04, 3.12$\pm$0.1, 0.98$\pm$0.05,  and 1.4$\pm$0.1\,$\times 
10^{-11}$\,erg\,cm$^{-2}$\,s$^{-1}$
during  the quiescent dips, the flares, the post-flare dips, and quiescence, respectively. These are 
consistent with those found in  2009 (dM10). 

\begin{figure}
\includegraphics[width=8.cm,height=\columnwidth,angle=-90]{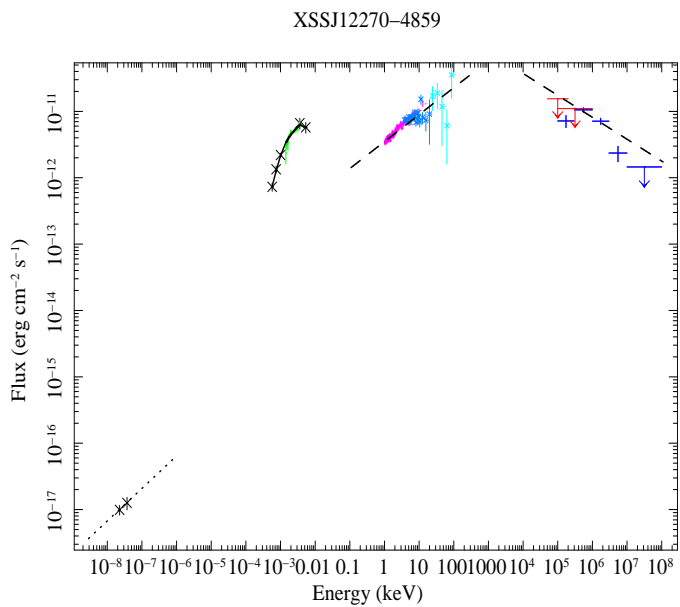}
\caption{Combined unabsorbed SED using the gamma-ray fluxes from \emph{Fermi}-LAT (blue) and 
\emph{AGILE}/GRID 
(red),  the \emph{XMM-Newton} (magenta), \emph{RXTE} (light blue), and \emph{INTEGRAL} (cyan)  
average X-ray 
spectrum, the \emph{XMM-Newton} OM UV and U band (black stars)
fluxes from dM10, the optical spectrum  from \citet{masetti06} (green), the 2MASS nIR measures and the ATCA radio 
measures from \cite{Hill} (black stars). Overlaid are the simple power-law forms  derived from the X-ray and 
gamma-ray spectra and from the radio-flux ratios as well as the best-fit composite blackbody model to 
the 
UV/optical/nIR fluxes. See electronic version for colour figure.}
\label{SED}
\end{figure}

\begin{figure}
\includegraphics[width=\columnwidth,height=7.cm,angle=0]{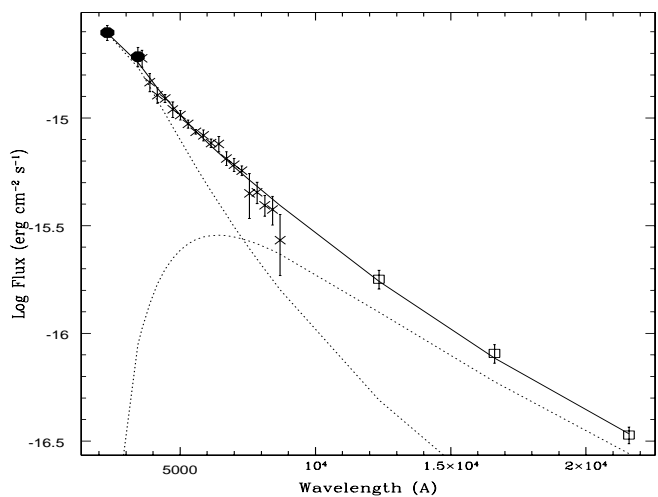}
\caption{Unabsorbed UV to nIR SED, obtained combining the \emph{XMM-Newton} OM UVM2 and U band fluxes 
(filled circles), the
optical spectrum from \citet{masetti06} (crosses), and the 2MASS fluxes (empty squares) together with 
the best-fit composite (solid line) function with  two blackbodies with temperatures of 12800\,K and 4600\,K (dotted 
lines).}
\label{SED_OPT}
\end{figure}

To construct the broad-band SED we implemented the X-ray unabsorbed spectrum obtained in dM10 using 
the \emph{XMM-Newton}, \emph{RXTE}, and 
\emph{INTEGRAL} data. This is   reported in Fig.\,\ref{SED} together with the best-fit power-law to 
the quiescent X-ray average 
spectrum.
We also overlaid the extinction-corrected \emph{XMM-Newton} OM UV and U band average fluxes (dM10), 
the extinction-corrected optical  spectrum by \cite{masetti06}, and the 2MASS nIR fluxes (see 
dM10 for details on 
extinction). The UV/optical and nIR data are consistent with each other, but are off by more than one 
order of 
magnitude from that extrapolating the X-ray power-law. The simultaneity of the UV and U 
band data with  the X-ray data then implies a different component contributing to  the UV/optical/nIR flux. 
We therefore analysed the UV/optical/nIR SED and found that it is best fitted by a composite function consisting 
of two blackbodies at $\rm T_h = 12800\pm 600$\,K and $\rm T_c = 4600\pm 250$\,K (uncertainty at 1$\sigma$ 
c.l.) ($\rm \chi^2/d.o.f.$=11/20). In Fig.\,\ref{SED_OPT} we report the UV/optical/nIR SED, where the 
optical spectrum from 
\citep{masetti06} was rebinned after removal of emission lines together with the two-component 
blackbody function. The latter  is also reported in Fig.\,\ref{SED}. The projected areas of the hot 
component  is 5$\%$ of the cool one. The low-temperature 
blackbody could be compatible with that of a late-type star in the range K2-K5\,V, which is of 
earlier type than
what is estimated in dM10, assuming an orbital period of 4.3\,h. For a K2-K5\,V star (log\,g=4.5), 
the projected area would imply 
a distance of 1.7$\pm$0.3\,kpc. The J and K band fluxes of this component are $98\%$ and $99\%$ the total observed fluxes, 
respectively. Hence, it is reasonable to assume that the nIR flux is almost totally dominated by the donor star.  We then 
also use the lowest J-band magnitude observed in dM10 photometry (J=16.9), corrected for interstellar 
absorption $\rm A_J=0.12\,mag$ (dM10), and the absolute magnitudes corresponding to a K2-K5\,V star 
\citep{knigge06}, thus obtaining 
a distance of 2.3--3.6\,kpc. If 
the secondary star is indeed of K-spectral type, it would in turn indicate a binary period 
in the range 7--9\,h \citep{Smith_Dhillon98}. These values and the  $\sim$6\,h variability detected in the near-UV might 
suggest a longer  orbital period. The high-temperature component instead suggests a non-stellar  
contribution from an  accretion disc or flow, because  the optical spectrum is characterized by emission 
lines. The projected area of the hot emitting region is $\rm \sim 7 \times 10^{20}\,d_{1\,kpc}^2 \,cm^{2}$.

Assuming the  association with the 2FGL source, we also constructed the gamma-ray portion of the 
spectrum using the 
\emph{AGILE}/GRID upper limits and the
 \emph{Fermi}-LAT fluxes as listed in the $\rm 2^{nd}$ source catalogues (Fig.\,\ref{SED}). The 
spectrum 
is also reported together  with the best-fit 2FGL catalogue power-law with photon index 2.33. The 
high-energy portion of the 
gamma-ray spectrum might suggest a decay that \citet{Hill} fitted with an exponential cut-off at 
$\sim$4.1\,GeV, 
found  to be significant at 4$\sigma$. The logParabola used in the 2FGL catalogue could also support this 
 feature. The \emph{AGILE}/GRID upper limits are roughly compatible with the  \emph{Fermi}-LAT measures.  
The gamma-ray SED  does not support a peak in the 1-100\,MeV range, as suggested in dM10.
Therefore there are no strong evidences that the X-ray and gamma-ray fluxes are linked. 

We furthermore include the  dual band ATCA fluxes  at 5.5\,GHz and 9\,GHz from \citet{Hill}, acquired 
in 2009,  of the radio counterpart to XSS\,J1227.  Although not 
contemporaneous, the X-ray and gamma-ray histories  allow us to adopt the radio measures for the broad-band 
SED. The flux-density ratio at these frequencies, assuming a power-law distribution $\rm S_{\nu} 
\propto \nu^{\alpha}$, gives 
$\alpha$ = -0.5 \citep{Hill}.
Taken at its face value,  it is consistent with optically thin synchrotron emission. An extrapolation of 
this power-law to higher energies 
falls well below the X-ray spectrum but roughly matches the  soft gamma-ray flux  at $\sim$ 100\,MeV. 
However, we note that the large uncertainties in the spectral index derived by \citet{Hill} may also allow 
a flat radio spectrum.

\section{Discussion}

We have presented  new X-ray data of XSS\,J1227 and new gamma-ray data of its positionally associated 
\emph{Fermi}-LAT  counterpart to infer the long-term temporal history  and the 
possible nature of this intriguing source.

\subsection{A steady source: the persistent, flaring, and dipping states }
 
XSS\,J1227 is found at about the same flux level ($\rm F_{2-10\,keV} \sim 1\times 
10^{-11}\,erg\,cm^{-2}\,s^{-1}$) over 7\,yr of observations,
displaying peculiar short-term behaviour with flares and dips in all observations. 
Flares are simultaneously 
detected in both X-rays and UV band on timescales of $\sim$600-850\,s.
All X-ray flares occur in the form of flare-dip pairs. In dM10 we drew comparison with 
type-II bursts sources,  although  with different flare energetics and duration. 
Flares appear to occur on a timescale of $\ga $9-11\,h and are grouped in multiple 
shorter events with a total duration of $\ga$1\,h.  Flares were also reported to occur in the nIR 
band by
\citet{Saitou11}. The large variability ($\sim$0.4\,mag) detected in our previous 
J-band data (dM10) was not recognized to be in the form of flares due to the poor statistics. 
The optical photometry by \citet{Pretorius09} and dM10 also shows large-amplitude short term 
variations  that could be affected by flares (dM10). 

While X-ray flares are characterized by unvariant spectral shape with respect to the quiescent persistent 
level, the 
post-flare dips involve absorbing material that is dense ($\rm N_H = 4\times 10^{21}\,cm^{-2}$),  
partially   ($\sim$60$\%$) covering the X-ray source. The  simultaneous  occurrence of flares at 
lower energies in the 
near-UV  corroborates the interpretation given in dM10, where flare-dip 
pair events are due to  accretion events onto the compact 
object  followed by a slow refilling of the missing regions of an accretion disc. The emptied region 
should be extended because it is detected  at UV and near-UV energies. 

On the other hand, the ordinary dips observed during quiescence in  X-rays and near-UV appear to 
occur randomly. Here we note that optical ground-based photometry
does not reveal such sharp dips (Pretorius 2009,dM10), indicating that dips originate close to the compact star.
 The sharpness of their profiles, the different lengths (from $\sim$15\,s up to 
$\sim$1100\,s), the timescales of decays/rises  and the  invariance (or possibly slight softening) 
of the X-ray  spectrum  might point to occultation due to  discrete material. If this matter is 
located at the disc edge, the 
dip ingresses/egresses might  be used to estimate the size of the eclipsed X-ray emitting region, 
similar to LMXB dippers. However,  LMXB dippers 
are known to show intensified dipping activity at specific orbital phases with  hardening of the
spectrum due to  absorbing material located close to the disc bulge \citep{Boirin05,Iaria07,DiazTrigo09}. 
The frequent and apparently non-periodic occurrence of deep dips in XSS\,J1227 could be related to opaque material 
distributed randomly above the disc. Although caution should be taken, we follow \citet{Church_Balucinska04} for LMXB 
dippers, using the longer dips where the  X-ray count rate drops to zero. The
range of ingress/egress times is  $\sim$40-90\,s. We assume an accretion disc radius  $\rm R_{disc}\sim$0.90\,$\rm 
R_{Lobe,1}$. Given 
 the uncertainty in the orbital period, we use the wide range $\sim$4-9\,h and  adopt 
a mass ratio q=0.8. With these values,   the mass of the companion star would be in the range 
0.4-1\,M$_{\odot}$
 \citep{Smith_Dhillon98} and the disc radius would result $\sim$ 4-5$\times 10^{10}$\,cm. 
If the material is located at this  distance and the angular size of the material is larger than 
the X-ray emitting accretion disc corona (ADC) (see Eq.\,1 of \citet{Church_Balucinska04}), the time 
of ingress is related to the radius
of the emitter. In this case, we obtain $\rm R_{ADC} \sim 2-8\times 10^8\,cm$. This is about one 
order of magnitude 
smaller than those  determined for LMXB dippers \citep{Church_Balucinska04}. The size of the ADC scales with the X-ray 
luminosity \citep{Church_Balucinska04}  would imply  $\rm L_X \sim 0.5-1.0\times 
10^{36}$\,erg\,s$^{-1}$. This is about two orders of magnitude larger than that derived for a 
distance of 1\,kpc (dM10). Even
at  distances of 2.4-3.6\,kpc, the X-ray (bolometric)  luminosity is lower (3-8$\times 
10^{34}$\,erg\,s$^{-1}$). Therefore, either the disc corona is 
limited  in extension or  the occulting material is not located at the adopted disc rim. 

In addition, the dips are observed in the UV,  implying that the hot 13\,kK component that 
dominates the UV light also suffers 
occultation. This component  has an emitting area of $\rm \pi (0.2\,R_{\odot})^2\,d_{1\,kpc}^2$. This is smaller 
than the adopted disc radius but larger than those
determined for similar hot components in LMXBs and attributed to the bright disc hot spots or a region closer to the 
centre \citep{Froning11,Hynes12}. The UV/optical emitting region is larger than that estimated  for 
the X-ray emitter. This
can  explain why the UV flux deficiency is lower ($\sim 40\%$)  than that in the 
X-ray band during the dips. 

Furthermore, the fact that the ratio of bolometric fluxes between the X-ray and UV components is 
$\sim$6 and  the X-ray peak intensity at flares is about twice that in the UV indicates strongly
that the UV light originates from X-ray reprocessing in a larger region surrounding the X-ray source.

In summary, the long-term X-ray history of XSS\,J1227 over 7\,yr and the gamma-ray light curve of the 
associated source 
2FGL\,J1227 over 4.7\,yr  indicate that this peculiar binary is a persistent source in both energy ranges and that  
it always displays  the same X-ray variability characteristics.  We cannot, however, exclude the 
possibility that the gamma rays are also emitted in the form of flares with shorter time scales than 
4\,d.

\subsection{ An LMXB hosting a MSP?}

Analogies  to the MSP binary PSR\,J1023+0038  \citep{Tam} were suggested by \citet{Hill}
because of similarities of the optical spectra (both sources were formerly proposed as CVs), 
the existence of  
radio and \emph{Fermi}-LAT
counterparts and the short, putative  $\sim$ 4\,h orbital period, and large optical flickering. 
PSR\,J1023+0038 was proposed to be  active in gamma rays when accretion switches off and evidence of 
accretion stopping at least once comes from an  optical, although non-simultaneous, spectrum  (see \citet{Tam} 
and reference therein). The possibility that accretion switches off in XSS\,J1227 was left open in 
\citet{Hill}, but
the present study strongly indicates that the source has always been in an 
accretion state. This is further confirmed by recently acquired optical spectroscopy 
(de Martino et al. in prep.).  

We draw a further comparison with MSP binaries  detected by \emph{Fermi}. The  gamma-ray spectra of  
PSR\,J1023+0038 and 2FGL\,J1227  are rather similar, with a 
steep power-law index $\Gamma$ 2.5 (in PSR\,J1023+0038) and 
$\Gamma$=2.3 in 2FGL\,J1227. Although not strongly constrained,  a power-law exponential cut-off 
model for PSR\,J1023+0038 gave  $\Gamma$=1.9  and $\rm  E_{cutoff}$=0.7\,GeV \citep{Tam}
as compared with $\Gamma$=2.2 and $\rm E_{cutoff}$=4.1\,GeV inferred for 2FGL\,J1227 by \citet{Hill}.
\citet{Ransom11}  fitted  the  gamma-ray spectra of PSR\,J0614-3329, PSR\,J1231-1411 and PSR\,J2214+3000, with 
a  power-law exponential cutoff, obtaining $\rm E_{cutoff}$ in the range 2.5--4.5\,GeV and $\Gamma\sim$1.4.  
 \citet{Kerr12} find for PSR\,J0101-6422 $\rm E_{cutoff}$=1.9\,GeV and $\Gamma$=0.9. Also 
for the very recently discovered black-widow MSP PSR\,J1816+4510, \citet{Kaplan12} found $\Gamma$=2 
and $\rm  E_{cutoff}\sim$7.5\,GeV.
Therefore, based on the gamma-ray spectrum, 2FGL\,J1227 is not much different from the
variety of MSP binaries recently discovered by \emph{Fermi}. 

The X-ray (0.2-100\,keV) to gamma-ray (0.1-100\,GeV) luminosity ratio of our source is $\sim$0.8, whilst most 
of the MSP in LMXBs have lower ratios  by 2--3 orders of magnitudes   
\citep{Bogdanov06,Ransom11,Bogdanov11}. Our  exploration of the overall SED also indicates that 
the X-ray  flux,  dominated by a non-thermal  component, is higher than that expected from 
rotation-power MSP emission. In the latter, the thermal flux from the NS atmosphere either dominates 
\citep{Ransom11} or is detectable, as in 
the case of PSR\,J1023+0038 with a thermal 
fraction  $\sim$3-6$\%$ \citep{Bogdanov11} (see also \citet{Takata12}).  The non-thermal X-ray  
emission, if  originating  from an intrabinary shock, such as in the black-widow MSPs,  
would be the extension  of the gamma-ray flux at these  energies \citep{Takata12}, and this is not 
the case for 
XSS\,J1227. Intrabinary shock as source of non-thermal emission in the quiescent MSP 
binary PSR\,J1023+0038 was  proposed by \citet{Bogdanov11}. 

Our search for X-ray msec pulses in XSS\,J1227 only provided upper limits $\sim 15\%$ to pulse 
amplitudes. In accreting MSPs, pulse amplitudes in excess of 10$\%$ are rarely observed,
but fractional amplitudes can range between 0.43$\%$  and  $\sim 30-40\%$   (see review by
\citet{PatrunoWatts}). Here we note that in PSR\,J1023+0038  the X-ray pulse amplitude was found to be
 $11\pm2\%$ \citep{Archibald10}. Therefore, we cannot exclude that  XSS\,J1227 
harbours a fast-spinning pulsar. However, in non-accreting MPS the X-ray pulses would arise from the 
NS atmosphere or hot 
spots that we have not detected in the X-ray spectrum of XSS\,J1227. A non-thermal component arising from intrabinary shock 
would only  produce an X-ray orbital modulation, as is the case of PSR\,J1023+0038, which  shows 
large amplitude orbital variability \citep{Bogdanov11}. While in the  
2009 \emph{XMM-Newton} data, a possible weak X-ray variability could be present, this is not detected 
in the  higher quality EPIC-pn data in 2011. Hence, all this points to an accretion-driven X-ray 
emission.

Accretion acting in this source is further corroborated by  the UV/optical/nIR portion of the SED, 
which reveals two spectral components: a hot one at  $\sim$13\,kK and a cool one at 4.6\,kK. A hot, 
but single, spectral 
component was also recently identified in the UV/optical SED of the MSP binary PSR\,J1816+4510 
\citep{Kaplan12}, making it the black-widow binary pulsar hosting the hottest low-mass companion 
($\ga$10\,kK). We however note  that no nIR data are available to date for PSR\,J1816+4510. In 
XSS\,J1227, the cool component is instead revealed from the nIR 2MASS measures and its temperature 
suggests the contribution from
a late-type K2-K5 companion,  earlier than that estimated in dM10. Hence, the UV/optical component 
can  be reasonably ascribed to accretion. The temperature is,  however, higher than that of accretion 
discs (see \citet{Hynes12}) and the size of the emitting 
region could suggest that it originates closer to the compact object. The X-ray and UV bolometric luminosity ratio is 
$\sim$6, compatible with an X-ray heated region. We note that this ratio has been proposed to be
a discriminant between NS and BH binaries, with NS having X-ray-to-UV flux ratios 
about ten times higher than BH systems \citep{Hynes12}.

 If the radio spectral power-law index is truly $\alpha= $-0.5, then the 
extrapolation to the gamma-ray regime roughly matches the observed flux at 
$\sim$100\,MeV.  However, searches for fast pulses in the radio counterpart of XSS\,J1227  failed  
\citep{Hill}. Therefore, while the flux ratio could be consistent with non-thermal emission, the lack 
of   detection of pulses does not favour a rotation-powered radio pulsar.

\begin{figure}
\includegraphics[width=9.cm,height=\columnwidth,angle=-90]{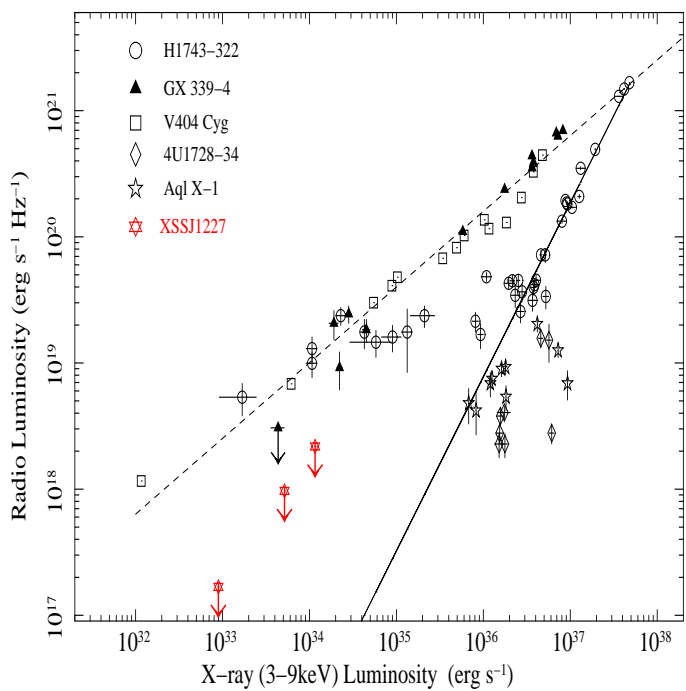}
\caption{Radio luminosity at $\sim$9\,GHz versus  3-9\,keV X-ray luminosity for the BH candidates  
GX\,339-4, 
H\,1743-322 and V404\,Cyg during their hard X-ray states, together with the atoll-type NS binaries Aql\,X-1 and 
4U\,1728-34
during their island states (see footnote\,2 for details on the data). The radio/X-ray luminosity correlation $\rm 
F_{radio} \propto F_{X}^b$ with b$\sim$0.6 for BH binaries and b$\sim$1.38 for island state NS binaries
is also reported. The positions of XSS\,J1227 at the three distances 1\,kpc, 2.4\,kpc, and 3.6\,kpc 
are marked in red.   Adapted from \citet{Coriat11}. See electronic version for colour figure.}
\label{CORIAT_FIGURE}
\end{figure}

\subsection{ An LMXB powering a jet?}

The hard X-ray spectrum and the low-luminosity of XSS\,J1227 indicate an LMXB in a deep faint hard 
state. 
From the detection of quasi-simultaneous X-ray and nIR flares
and the low X-ray luminosity,  \citet{Saitou11} argued that XSS\,J1227 could be a microquasar with a 
synchrotron jet in a prolonged low-luminosity state
at $\rm 10^{-4}\,L_{Edd}$ for a stellar mass BH  or NS at 1\,kpc. 
As a consequence  of synchrotron cooling in an optically thin medium, the energy break expected in 
these binaries is at 
$\sim$500\,keV \citep{Saitou11}, much lower than that detected at $\sim$4\,GeV in 2FGL\,J1227.
 
High-energy gamma rays were detected in galactic binaries hosting high-mass early-type 
 stars, such as LS\,5039, LSI +$\rm 61^{\circ}303$ \citep{abdo09a,abdo09b,Hadasch}, 
1FGL\,J1018.6+5856 \citep{Ackermann12}, and 
Cyg\,X-3 during  flares   \citep{Tavani09,abdo09c}. The latter is the only gamma-ray source  with a 
BH candidate 
while the others are proposed to possibly harbour NSs. The gamma-ray spectrum of LSI+$\rm 
61^{\circ}303$ is 
compatible with 
a power-law exponential cut-off at  $\sim$4\,GeV \citep{abdo09a,Hadasch},  
while LS\,5039 shows a power-law exponential cut-off at $\sim$2\,GeV \citep{abdo09b,Hadasch}. The nature of 
GeV emission from these binaries is still unclear and controversial \citep{Hadasch,Papitto12,Torres12}.  
 They could be 
powered either by  magnetospheric emission, as seen in many gamma-ray detected pulsars, or by 
electrons in  the jets  accelerated up to TeV energies (leptonic model) \citep{Hadasch}.  For 
Cyg\,X-3, the
Fermi flux  is compatible with the extrapolation of the X-ray tail to 100 MeV and hence is believed 
to be the high-energy tail 
of the soft gamma-ray spectral component. Also, in this case the origin of gamma-ray photons is 
unclear and is proposed to be  leptonic \citep{abdo09c}.
Since no LMXB with a radio jet  has been detected in the gamma rays so far, it is difficult to predict the origin of 
high-energy  gamma-ray emission.

We alternatively explored the radio-to-X-ray emission from LMXBs hosting NS and BH in the hard faint quiescent 
states. The persistent radio emission from BH binaries during these states is generally consistent 
with a flat or 
slightly inverted spectrum that, in  analogy with compact extragalactic radio sources, is attributed  
to the 
superimposition of a number of peaked synchrotron spectra  generated in a jet \citep{Gallo10}. Atoll NS 
binaries in low-hard states 
(island states) have also been detected at radio frequencies, though most are upper limits with a 
handful of systems detected simultaneously with the X-rays \citep{MigliariFender06}. Usually, 
dual-band radio spectra 
are not well constrained and could be consistent with  flat ($\alpha \simeq $ 0) or optically thin 
$\alpha \sim -0.6$ slopes, such as Aql X-1 or 4U\,1728-34 \citep{Migliari03,MigliariFender06}. 
In NS binaries, jets can nevertheless be powered during these low states but are about one order
of magnitude less powerful than BH jets. The radio spectrum of XSS\,J1227 is not well constrained,  
$\alpha$=0.5$\pm$0.6 \citep{Hill}, leaving the possibility of a similar origin as those in BH or NS binaries.

The BH binaries spend most of their time in such states, which could be compared to the prolonged 
low-state of XSS\,J1227. The X-ray emission from these binaries originates in an optically thick accretion disc with 
a hard component, which could be an ADAF-like (Advection Dominated Accretion Flow (ADAF) 
\citep{Narayan94} ) or 
from a jet.  The  jets are coupled with the accretion flow, although the disc-jet connection is still 
unclear. The radio emission from the jet was found to be correlated with that in the X-rays during the  hard state: 
$\rm F_{radio} 
\propto  F_{X}^b$, with b$\sim$0.5-0.7 \citep{Gallo03}. A similar relation was found  for atoll
NS binaries in the island states but with a steeper  correlation coefficient b$\sim$1.4 
\citep{MigliariFender06}. 
 Hence, the  X-ray emission  in the BH systems is radiatively inefficient, with an  approximate 
relation $\rm L_X \propto \dot M^{2}$, whilst the jet power in both classes scales linearly with the accretion 
rate. Keeping in 
mind the large uncertainty in the radio spectral slope, we have explored the possibility that XSS\,J1227 could 
host a compact jet following \cite{Coriat11}, where the radio/X-ray luminosities of BH binaries and 
NS LMXBs are compared. 
This is depicted
in Fig.~\ref{CORIAT_FIGURE},  adapted from Fig.\,5 of \cite{Coriat11}. Here, the radio (9\,GHz) 
luminosity against the 
X-ray  3-9\,keV 
unabsorbed luminosity is shown for a sample of BH candidates in  LMXBs (GX\,339-4, H\,1743-322 and 
V404\,Cyg) during hard 
X-ray states and for the two atoll NS binaries in the island state, Aql X-1 and  
4U\,1728-34~\footnote{The radio-vs-X-ray 
luminosity diagram is constructed using data for  GX\,339-4 from  \citet{Corbel03} and a distance of 8\,kpc, for  
H\,1743-322  from 
Table\,1 of   \citet{Coriat11} and distance of 8\,kpc, and for V404\,Cyg from \citet{Corbel08} and a 
distance 
2.39\,kpc. For the atoll NS binary Aql X-1, we use data from \citet{Miller-Jones10} and distance 
5\,kpc, and for 4U\,1728-34
we use data from \citet{MigliariFender06} and distance 4.6\,kpc.}  In the radio vs. X-ray 
luminosity  diagram, the BH candidate H\,1743-322 moves to lower X-ray luminosities,  transiting the 
two 
correlations and  joining the standard BH correlation at low luminosities \citep{Coriat11}. 
This is the first BH candidate to follow both correlations depending on the X-ray luminosity. 

We locate XSS\,J1227 in this diagram by using the ATCA 
9\,GHz and 3-9\,keV fluxes adopting the three values for the distance, 1\,kpc, 2.4 \,kpc, and 
3.6\,kpc.  For the sake 
of caution,  these are marked as upper limits because of the uncertainties in the radio slope and 
hence in the 
contribution of a jet.
XSS\,J1227  appears  to be rather away from the NS binaries locus and, in case of a jet in this 
system, it is closer 
to the BH correlation. Its  position could move upwards, matching at 10\,kpc the transition region of  
H\,1743-322, where the latter joins the standard BH correlation. 
While we cannot infer the true nature of the compact star in XSS\,J1227 based on this diagram alone,  
if a jet is present, it could possibly favour a BH 
LMXB. If so and similarly for H\,1743-322 \citep{Coriat11}, this binary would be in an extremely  
deep  low hard state, 
where the  accretion  flow  becomes radiatively inefficient. This flow is then 
dominated by advection in 
which a  significant fraction of the energy is advected and not radiated away \citep{Narayan94}. The energy 
either crosses 
the BH event 
horizon or is expelled in outflows. Systems dominated by jet emission, where most of the energy is channeled in a jet, emit 
X-rays at the base of the jet.  These are produced either by synchrotron emission or  by inverse Compton scattering by 
outflowing particles. 
In the previous section, we derived an approximate estimate of the size of the X-ray emitting region, 
which is much smaller  than that of typical ADC in LMXBs. 
\citet{Coriat11} also suggest that H\,1743-322  possesses  two components, one radiatively efficient 
and one inefficient, which co-exist and dominate alternatively in the X-ray band during the hard 
state. In the case of XSS\,J1227, the bulk of X-rays does not seem to originate in a jet 
 because of the evidence of accretion.
Further radio data would be important to assess the  true  nature of the radio emission.

\section{Conclusions}
 
We have presented new X-ray data to study the long-term history of XSS\,J1227 and to infer whether 
the X-ray emission 
is variable on timescales down to msec in order to shed light on the nature of this peculiar source. We here summarize the 
main conclusions:\\

\begin{itemize}
\item
  The long-term X-ray history of XSS\,J1227 over a time-span of 7\,yr shows that it is a persistent low 
X-ray luminosity source and not a transient one. 
The associated source 2FGL\,J1227 is also found to be a steady 
high-energy gamma-ray 
source over an overlapping period of 4.7\,yr.  The co-existence of X-ray and gamma-ray
 emissions does not favour a MSP binary nature if the two sources are associated.\\
\item
 XSS\,J1227 displays flares and dips in all of the X-ray observations, with flares possibly occurring on a 
timescale of 9-11\,h and duration of $\sim$ 1\,h. Flares are grouped in multiple events. We confirm previous finding 
of the occurrence of flare-dip pairs. The  spectral hardening in these dips are due to an
intervening dense, $\rm N_H \sim 4\times 10^{21}\,cm^{-2}$, absorber covering $\sim 60\%$  of the 
X-ray source.
Intense dipping activity is found during quiescence. These dips have variable 
lengths from $\sim$10\,s to 1100\,s. A marginal spectral softening is detected in the longer dips. We estimate the size of 
the X-ray emitting region to be $\sim 2-8\times 10^{8}$\,cm, much smaller than those derived for LMXB 
ADC dippers. The UV flares and dips are found to occur simultaneously with the X-ray  ones.\\

\item
 We  searched for fast pulsations down to msec in the fast-timing X-ray data obtaining upper limits 
to fractional pulse 
amplitudes of 15-25$\%$. These do not exclude an MSP in this system. A rotation-powered MSP 
could be favoured by the 
SED at  gamma-ray energies and possibly at radio frequencies, but not from the X-ray and 
UV/optical/nIR  data. The X-ray source is consistent with being an accretion-powered binary.\\

\item
 From the combined UV/optical/nIR spectrum, we find evidence of a hot component at $\sim$13\,kK and a 
cool one 
at $\sim$4.6\,kK. The former is consistent with an X-ray-heated accretion region, which is smaller 
than that of a 
tidally truncated accretion disc but larger than those  found in other LMXBs with either NS or BH. The 
cool component
would suggest  a late-type K2-K5 companion star and in turn a distance of 2.4--3.6\,kpc.
Then the  orbital period would be in the range of 7--9\,h. This would embrace  the 4.3\,h periodicity previously inferred 
(dM10) if it 
were the first harmonic. A 20$\%$ variability at 6.4\,h is found in the quiescent near-UV photometry, 
further suggesting a long orbital period. \\
\item
 The X-ray-to-UV luminosity ratio is  compatible with an LMXB hosting an NS that is much larger than 
that 
observed in BH LMXBs.  On the other hand, the uncertainties in the radio spectral slope may also allow the 
possibility of a compact jet. In this case,  the radio and X-ray 
luminosities might indicate XSS\,J1227 as a BH binary rather than an NS atoll system during island 
states. The 
persistent low 
luminosity of this source could be due to a radiatively inefficient accretion flow, where the radio
emission originates in a jet but the X-ray emission does not. If a BH binary,  XSS\,J1227 were the 
first LMXB to be
associated with a high-energy gamma-ray source. 
\end{itemize}

To further progress on the intriguing nature of this source,  it is crucial both to obtain new 
observations in order to infer whether flares also occur in 
 the radio domain and to study the radio spectrum and optical  spectroscopy in order to determine the 
true orbital period.

\begin{acknowledgements}
DdM and TB acknowledge financial support from ASI under contract 
ASI/INAF I/009/10/0. AP acknowledges the support of the grants AYA2012-39303,
SGR2009- 811,  and iLINK2011-0303.
This research has received funding from the European Commnunity's Seven Framework
Programme (FP7/2007-2013) under grant agreement ITN 215212.
 This work made use of data supplied by the UK Swift Science Data Centre 
at the University of Leicester. DdM wishes
 to thank Dr. Norbert Schartel and the ESAC staff for their help in 
obtaining  the \emph{XMM-Newton} data.

\end{acknowledgements}

\bibliographystyle{aa}
\bibliography{arXivNEW}

\onecolumn
\begin{longtab}
\begin{longtable}{l l l l r c}
\caption{\label{observ} Summary of the observations of XSS\,J1227}\\
\hline\hline
Telescope  & Instrument	& Date & UT (start) & Exposure time\,(s) & Net count rate($\ctss$)\\ 
Prog. ID   &               &                       &  &                   &                         \\
\hline
\endfirsthead
\hline
\caption{continued.}\\
\hline\hline
Telescope & Instrument & Date  & UT (start) & Exposure time\,(s) & Net count rate ($\ctss$)\\
Prog. ID  &    &     &               &                   &                         \\
\hline
\endhead
\hline
\endfoot
& & & & & \\ 
 \emph{XMM-Newton}~\footnote{Total EPIC count rates in the 0.2-10\,keV range}  & EPIC-pn		& 2011-01-01    & 07:06			& 30\,045  & $3.71\pm 0.09$ \\
 65678      	   & EPIC-MOS		&		& 06:47			& 31\,022  & $1.09\pm 0.01$ \\
		& OM-U 		  	&		& 06:56		& 2\,699   & $5.19 \pm 0.04$	\\
		&  		  	&		& 07:46		& 2\,700   &  	\\
		&  		  	&		& 09:07		& 2\,700   &  	\\
		&  		  	&		& 09:57		& 2\,699   &  	\\
		&  		  	&		& 10:48		& 2\,700   &  	\\
		&  		  	&		& 11:38		& 2\,499   &  	\\
		&  		  	&		& 12:25		& 2\,499   &  	\\
		&  		  	&		& 13:12		& 2\,499   &  	\\
		&  		  	&		& 13:59		& 2\,498   &  	\\
		&  		  	&		& 14:46		& 2\,500   &  	\\
& & & & & \\
\emph{RXTE}
~\footnote{Total \emph{RXTE}/PCA count rates in the 2-9\,keV range.}    & PCA  & 2011-01-03 & 00:27 & 2013 & 1.72$\pm$0.06\\
96309                      &      & 2011-01-10 & 14:11 & 2000 & 1.83$\pm$0.06 \\
                           &      & 2011-01-16 & 17:38 & 1917 & 1.94$\pm$0.05\\
                           &      & 2011-01-23 & 12:37 & 2082 & 2.21$\pm$0.07\\
                           &      & 2011-01-31 & 13:26 & 1919 & 2.03$\pm$0.07\\ 
                           &      & 2011-02-07 & 13:38 & 2052 & 0.96$\pm$0.06\\ 
                           &      & 2011-02-14 & 15:05 & 1947 & 1.84$\pm$0.06\\ 
                           &      & 2011-02-21 & 17:49 & 1836 & 0.97$\pm$0.06\\ 
                           &      & 2011-02-28 & 12:51 & 2476 & 1.59$\pm$0.06\\ 
                           &      & 2011-03-07 & 10:58 & 1977 & 2.44$\pm$0.07\\ 
                           &      & 2011-03-14 & 05:59 & 2043 & 2.05$\pm$0.07\\ 
                           &      & 2011-03-21 & 12:32 & 1962 & 0.89$\pm$0.06\\ 
                           &      & 2011-03-28 & 21:22 & 1874 & 0.56$\pm$0.06\\ 
                           &      & 2011-04-03 & 18:31 & 1794 & 1.07$\pm$0.06\\ 
                           &      & 2011-04-13 & 09:03 & 1919 & 1.73$\pm$0.06\\ 
                           &      & 2011-04-18 & 14:18 & 1770 & 2.25$\pm$0.07\\ 
                           &      & 2011-04-25 & 06:12 & 1962 & 1.90$\pm$0.06\\ 
                           &      & 2011-05-03 & 19:58 & 1586 & 1.05$\pm$0.08\\ 
                           &      & 2011-05-10 & 21:05 & 1922 & 1.82$\pm$0.07\\ 
                           &      & 2011-05-16 & 13:17 & 1909 & 1.03$\pm$0.06\\ 
                           &      & 2011-05-23 & 17:37 & 2010 & 1.63$\pm$0.06\\ 
                           &      & 2011-05-30 & 15:49 & 2034 & 0.91$\pm$0.06\\ 
                           &      & 2011-06-06 & 06:14 & 1742 & 2.06$\pm$0.07\\ 
                           &      & 2011-06-13 & 02:34 & 1807 & 1.09$\pm$0.06\\ 
                           &      & 2011-06-19 & 04:14 & 2043 & 2.27$\pm$0.07\\ 
                           &      & 2011-06-27 & 17:23 & 2099 & 1.83$\pm$0.07\\ 
                           &      & 2011-07-04 & 14:18 & 1824 & 1.34$\pm$0.06\\ 
                           &      & 2011-07-11 & 10:34 & 1789 & 1.06$\pm$0.07\\ 
                           &      & 2011-07-18 & 13:33 & 1965 & 1.51$\pm$0.07\\ 
                           &      & 2011-07-25 & 11:39 & 1716 & 1.51$\pm$0.06\\ 
                           &      & 2011-08-01 & 03:25 & 1947 & 2.06$\pm$0.07\\ 
                           &      & 2011-08-08 & 06:32 & 2102 & 1.94$\pm$0.06\\ 
                           &      & 2011-08-15 & 06:00 & 2004 & 0.84$\pm$0.06\\ 
                           &      & 2011-08-22 & 08:38 & 1756 & 0.90$\pm$0.07\\ 
                           &      & 2011-08-29 & 17:41 & 2132 & 1.61$\pm$0.06\\ 
                           &      & 2011-09-19 & 21:22 & 1859 & 1.10$\pm$0.06\\ 
                           &      & 2011-09-27 & 04:51 & 1656 & 0.83$\pm$0.06\\ 
                           &      & 2011-10-08 & 12:01 & 1731 & 1.03$\pm$0.07\\ 
                           &      & 2011-10-21 & 20:53 & 2025 & 1.03$\pm$0.06\\ 
                           &      & 2011-11-10 & 18:09 & 1933 & 2.24$\pm$0.07\\ 
                           &      & 2011-11-20 & 14:50 & 1860 & 0.80$\pm$0.06\\ 
                           &      & 2011-11-30 & 18:27 & 2094 & 0.84$\pm$0.07\\ 
                           &      & 2011-12-10 & 00:03 & 1787 & 0.84$\pm$0.06\\ 
                           &      & 2011-12-23 & 02:12 & 1947 & 1.04$\pm$0.06\\ 

& & & & & \\
\emph{Swift}~\footnote{Total \emph{Swift}/XRT count rates in the 0.3-10\,keV range.}  & XRT  & 2011-03-23 & 04:35 &  672 & 0.17$\pm$0.09\\
41135                      &      & 2011-03-23 & 06:11 &  732 & 0.26$\pm$0.11 \\
                           &      & 2011-03-23 & 07:47 &  701 & 0.26$\pm$0.09 \\
                           &      & 2011-03-23 & 09:24 &  729 & 0.28$\pm$0.12 \\
                           &      & 2011-03-23 & 11:00 &  729 & 0.35$\pm$0.12 \\
                           &      & 2011-03-23 & 12:36 &  852 & 0.27$\pm$0.09 \\
                           &      & 2011-03-23 & 14:13 &  792 & 0.30$\pm$0.09 \\
                           &      & 2011-05-10 & 16:32 &  634 & 0.67$\pm$0.15 \\
                           &      & 2011-09-15 & 07:55 &  854 & 0.32$\pm$0.13 \\
                           &      & 2011-09-18 & 09:48 &  809 & 0.07$\pm$0.06 \\
                           &      & 2011-09-19 & 04:59 & 1483 & 0.39$\pm$0.15 \\
                           &      & 2011-09-19 & 06:34 &  859 & 0.17$\pm$0.09 \\
                           &      & 2011-09-19 & 08:12 &  802 & 0.22$\pm$0.10 \\
                           &      & 2011-09-19 & 22:39 &  392 & 0.34$\pm$0.14 \\
                           &      & 2011-09-21 & 02:02 &  654 & 0.31$\pm$0.13 \\
                           &      & 2011-09-21 & 10:11 &  242 & 0.15$\pm$0.08 \\
                           &      & 2011-09-22 & 21:16 &  557 & 0.28$\pm$0.12 \\
                           &      & 2011-09-25 & 02:11 & 1293 & 0.39$\pm$0.15 \\
                           &      & 2011-09-26 & 02:20 &  557 & 0.24$\pm$0.09 \\
& & & & & \\
\emph{AGILE}~\footnote{\emph{AGILE}/GRID Pointing and spinning mode data: the 
start and end time and total exposure are reported.} 
   & GRID  & 2007-10-01 & 2009-10-31  & 5.3$\times 10^{6}$   &   \\
70                         &      & 2009-11-15 & 2011-05-15   & 3.8$\times 10^{6}$ & \\
\end{longtable}
\end{longtab}

\end{document}